\documentclass[sigconf,dvipsnames]{acmart}
\usepackage[utf8]{inputenc}

\PassOptionsToPackage{dvipsnames}{xcolor}
\PassOptionsToPackage{hyphens}{url}

\let\proof\relax
\let\endproof\relax

\usepackage{adjustbox}
\usepackage{amsmath}
\usepackage{amsthm}
\usepackage{anyfontsize}
\usepackage{booktabs}
\usepackage{caption}
\usepackage{color}
\usepackage{colortbl}
\usepackage{hyperref} 
\usepackage[capitalize]{cleveref}
\usepackage{cellspace}
\usepackage{circledsteps}
\usepackage{csquotes}
\usepackage{changepage} 
\usepackage{dashbox}

\usepackage{enumitem}
\usepackage{framed}
\usepackage{float}
\usepackage[scaled=0.95]{helvet}
\usepackage{mathtools}
\usepackage{microtype}
\usepackage{multicol}
\usepackage{multirow}
\usepackage{newfloat}
\usepackage{pifont} 
\usepackage{scalefnt}
\usepackage{setspace}
\usepackage[utf8]{inputenc}
\usepackage[binary-units=true,group-digits=integer,group-minimum-digits=4,group-separator={{,}},per-mode=symbol,detect-all]{siunitx}
\usepackage{tabu}
\usepackage{stmaryrd}
\usepackage{tcolorbox}
\usepackage{tabularx}
\usepackage{tikz}
\usepackage{thm-restate}
\usepackage{url}
\usepackage{xspace}
\usepackage{tcolorbox}
\usepackage{wrapfig}
\usepackage{xcolor}
\usepackage{subcaption}
\usepackage[normalem]{ulem}
\hypersetup{
     colorlinks=true,
     citecolor=blue,
     linkcolor=black,
}

\newif\ifDraft

\Draftfalse

\usepackage[lambda,adversary,asymptotics,sets,landau,probability,operators]{cryptocode}

\makeatletter
\g@addto@macro{\UrlBreaks}{\UrlOrds}
\makeatother

\usepackage[lambda,adversary,asymptotics,sets,landau,probability,operators]{cryptocode}
\usepackage{cleveref}

\theoremstyle{definition}

\bgroup\fboxsep=0pt

\NewEnviron{dottedbox}{
\par
\begin{tikzpicture}
\node[rectangle,minimum width=0.9\textwidth] (m) {\begin{minipage}{\textwidth}\BODY\end{minipage}};
\draw[dashed] (m.south west) rectangle (m.north east);
\end{tikzpicture}
}

\DeclareCaptionType[within=none,fileext=los,placement={!ht}]{construction}
\crefformat{construction}{construction~#2#1#3}
\Crefformat{construction}{Construction~#2#1#3}

\renewcommand{\pclnseparator}{:}
\renewcommand{\pcrlnseparator}{}
\renewcommand{\pclnspace}{0.4em}
\renewcommand{\pclnrspace}{0.4em}
\renewcommand{\pclnstyle}[1]{\color{gray}\text{\scriptsize#1}}
\definecolor{gamechangecolor}{gray}{0.90}

\tcbuselibrary{theorems}
\tcbuselibrary{skins}
\makeatletter
\crefformat{tcb@cnt@protocol}{protocol~#2#1#3}
\Crefformat{tcb@cnt@Protocol}{Protocol~#2#1#3}

\makeatother
\newtcbtheorem[auto counter]{protocol}{Protocol}{
	left=8pt,
	right=8pt,
	top=0pt,
	bottom=0pt,
    fonttitle=\normalsize,
    fontupper=\footnotesize,
	coltitle=black!,
	colbacktitle=black!0,
	colback=black!0, 
	boxrule=0.5pt,
	colframe=black!100,
	coltext=black!100,
	arc=0pt
}{proto}

\makeatother
\newtcbtheorem[auto counter]{functionality}{Functionality}{
	left=8pt,
	right=8pt,
	top=0pt,
	bottom=0pt,
    fonttitle=\bfseries\normalsize,
    fontupper=\small,
	coltitle=black!,
	colbacktitle=black!0,
	colback=black!0, 
	boxrule=0.5pt,
	colframe=black!100,
	coltext=black!100,
	arc=0pt
}{func}

\crefformat{tcb@cnt@functionality}{functionality~#2#1#3}
\Crefformat{tcb@cnt@Functionality}{Functionality~#2#1#3}

\newlist{algthenenumerate}{enumerate}{3}
\setlist[algthenenumerate]{label=\color{black}\arabic*:, font=\footnotesize, leftmargin=0.57cm, noitemsep,topsep=2pt,parsep=2pt,partopsep=1pt}

\newlist{algenumerate}{enumerate}{3}
\setlist[algenumerate]{label=\color{black}\arabic*:, font=\footnotesize, leftmargin=0.45cm, noitemsep,topsep=2pt,parsep=2pt,partopsep=1pt}

\usepackage{seqsplit}
\newcommand{\ttt}[1]{%
  \begingroup
    \protect\renewcommand{\seqinsert}{\ifmmode\allowbreak\else\-\fi}%
    \protect\texttt{\protect\seqinsert{\protect\seqsplit{\small#1}}}%
  \endgroup
}

\newcommand{\fb}{Flashbots\xspace}

\newtheorem{goal}{Goal}

\newcommand{\myparagraph}[1]{\vspace{0.25em}\noindent\textbf{#1.}}

\let\paragraph=\myparagraph

\newcommand*{\step}[1]{\Circled[fill color=black,inner color=white]{#1}}

\newcommand{\eg}{e.g.,~}
\newcommand{\ie}{i.e.,~}

\theoremstyle{definition}
\newtheorem{defn}{Definition}

\newcommand{\cameraAdd}[1]{%
  \ifDraft{\color{Blue}{#1}}\else#1\fi\xspace
}

\newcommand{\cameraDel}[1]{%
  \ifDraft{\sout{\color{Red}{#1}}}\else\fi\xspace
}

\copyrightyear{2022}
\acmYear{2022}
\setcopyright{acmcopyright}\acmConference[IMC '22]{Proceedings of the 22nd ACM
Internet Measurement Conference}{October 25--27, 2022}{Nice, France}
\acmBooktitle{Proceedings of the 22nd ACM Internet Measurement Conference (IMC
'22), October 25--27, 2022, Nice, France}
\acmPrice{15.00}
\acmDOI{10.1145/3517745.3561448}
\acmISBN{978-1-4503-9259-4/22/10}

\ccsdesc[300]{Security and privacy~Domain-specific security and privacy architectures}
\ccsdesc[100]{Security and privacy~Security protocols}

\begin{document}

\title{A Flash(bot) in the Pan: Measuring Maximal Extractable Value in Private Pools}

\author{Ben Weintraub}
\authornote{Both authors contributed equally to the paper.}
\affiliation{%
  \institution{Northeastern University}
  \city{}
  \state{}
  \country{}
}
\email{weintraub.b@northeastern.edu}

\author{Christof Ferreira Torres}
\authornotemark[1]
\affiliation{%
  \institution{SnT, University of Luxembourg}
  \city{}
  \country{}
}
\email{christof.torres@uni.lu}

\author{Cristina Nita-Rotaru}
\affiliation{%
  \institution{Northeastern University}
  \city{}
  \state{}
  \country{}
}
\email{c.nitarotaru@northeastern.edu}

\author{Radu State}
\affiliation{%
  \institution{SnT, University of Luxembourg}
  \city{}
  \country{}
}
\email{radu.state@uni.lu}

\begin{abstract}

  The rise of Ethereum has lead to a flourishing decentralized marketplace
  that has, unfortunately, fallen victim to frontrunning and
  Maximal Extractable Value (MEV) activities, where savvy participants 
  game transaction orderings \emph{within a block} for profit.
  One popular solution to address such behavior is \fb, a private pool with
  infrastructure and design goals aimed at eliminating the negative
  externalities associated with MEV.
  While \fb has established laudable goals to
  address MEV behavior, no evidence has been provided to show that these
  goals are achieved in practice.
   
  In this paper, we measure the popularity of \fb and evaluate if it is
  meeting its chartered goals. We find that
  (1) \fb miners account for over \SI{99.9}{\percent} of the hashing power in
  the Ethereum network,
  (2) powerful miners are making more than $2\times$ what they were making
  prior to using \fb, while non-miners' slice of the pie has shrunk
  commensurately,
  (3) mining is just as centralized as it was prior to \fb with
  more than \SI{90}{\percent} of \fb blocks coming from just two miners, and
  (4) while more than \SI{80}{\percent} of MEV extraction in Ethereum is
  happening through \fb, \SI{13.2}{\percent} is coming from other private pools.

\end{abstract}

\maketitle

\section{Introduction}
Many investors find the promise of Decentralized Finance (DeFi) appealing. This
has lead to cryptocurrency exchanges seeing over \num{78} billion USD in daily trade
volume~\cite{coinmarketcap}. Meanwhile, cryptocurrency donations to
Ukraine surpassed \num{63} million USD between February 26th and March 11th,
2022\cameraAdd{~\cite{ukraineCrypto}}, indicating the usefulness of cryptocurrencies for both private investors
and nation states.

Unfortunately, the exchanges in which these currencies are traded, and
especially those running on the Ethereum blockchain~\cite{ethereum}, are victim
to a costly type of malicious behavior: \emph{frontrunning}. Frontrunning is a concept
borrowed from traditional finance and popularized by Michael Lewis's bestselling
\emph{Flashboys: A Wall Street Revolt}~\cite{flashboys}. Frontrunning is a type of high-frequency trading
where one party uses privileged access to pools of pending trades to execute
their own trade before a targeted pending trade. There are several types of frontrunning, but all
rely on extracting profit from slippage caused by a publicly viewable
transaction.

\citet{daian2020flash} found, in 2020, that frontrunning analogous to
traditional markets is plaguing cryptocurrency exchanges: particularly
decentralized exchanges on Ethereum.
Frontrunning in the cryptocurrency world is also known as \emph{Maximal
  Extractable Value} (MEV); it occurs when a malicious peer on the blockchain
network learns of an uncommitted transaction and is able to exert some
control of the transaction ordering for its own
profit~\cite{daian2020flash}.
MEV is a widespread problem. \citet{ferreira2021frontrunner} found that
\num{199724} frontruns occurred from July 30th, 2015 to November 21th, 2020, for
a cumulative profit of \SI{18.4}{million\ USD}.

Since the pioneering work of \citet{daian2020flash}, the research community has
published numerous solutions for addressing this
problem~\cite{kelkar2020order,kelkar2021themis,kursawe2020wendy,chainlinkblog}.
However, the most popular solution, in practice, is called \emph{\fb}~\cite{flashbotsdocs}.
\fb works by creating a private transaction pool where \cameraAdd{non-mining participants (called
\emph{searchers} in \fb parlance) submit immutable bundles of transactions} to relays who broadcast
them to participating miners. \cameraAdd{The miners then mine the most
  profitable bundles with each block.} 
 \cameraAdd{Searchers must pay miners a fee in exchange for including their
   transactions in blocks. \fb is not strictly a defense against MEV. It allows
   transaction submitters to control transaction ordering, which means that
   \emph{anyone} can now be an MEV extractor. The only defense it offers is that
   submitters who do not want to be frontrun can also submit their transactions through \fb.
   This offers them protection, because the transaction ordering is guaranteed
   by \fb.}

Despite a number of structural flaws, \fb adoption
skyrocketed over the course of 2021. To date, however, this solution has not
been audited by external researchers.
In addition to \fb, there are other private transaction pools. The relationship
of these pools (and users thereof) to \fb is, to our knowledge, unexplored in
the literature.




In this paper we measure and analyze the impact of \fb, and evaluate whether or
not it is meeting its claimed goals. We also evaluate non-\fb private pools with
a focus on MEV extraction occurring in those pools. In particular, we highlight
three key findings.

$\bigstar$ \fb usage grew rapidly at first, and by now has captured nearly all
of the Ethereum network's hashing power. 
\cameraAdd{Only five months after its release, \fb had already reached a hash
  rate of \SI{97.6}{\percent} and was hovering around \SI{99.9}{\percent} after
  only eight months.}

$\bigstar$ \fb is disproportionately beneficial for miners at the expense of
non-miners. In addition, miners using \fb are making more profit than before
using \fb, while non-miners are making less. \cameraAdd{This has impacted usage, which has driven an exodus from the system by some users.}

$\bigstar$ Not all MEV extractors are using \fb. There is a small but powerful
contingent using other private pools. In some cases, the pools consist of a
single miner.

The rest of the paper proceeds as follows. We first give necessary background on
DeFi and the affected exchanges (\Cref{sec:bkg}). Next, we describe the
data we collected and analyzed (\Cref{sec:datasets}). We then move onto a
discussion on the usage of \fb (\Cref{sec:usage}) and an audit on how well it is
achieving its goals (\Cref{sec:audit}). In \Cref{sec:private}, we describe and
analyze other private pools in Ethereum, and how they are involved in MEV,
followed by related work in \Cref{sec:related_work}. We
conclude in \Cref{sec:conclusion} with a discussion on the efficacy of \fb as an
MEV solution, and whether or not it is a viable long-term solution.







\section{Background}\label{sec:bkg}
\subsection{Ethereum}

Ethereum~\cite{ethereum} is a decentralized blockchain that facilitates
cryptocurrency transactions and \emph{smart contract} execution. The blockchain
consists of blocks, which themselves consist of \emph{transactions}. Ethereum
nodes propagate transactions through the network where \emph{miners} collate
them into blocks. Transactions received over the network are collected locally
at each participating node in the \emph{mempool}. The mempool has no
blockchain-like guarantees of consistency.

Miners select a subset of the mempool transactions and execute a
\emph{proof-of-work} challenge~\cite{bitcoin}. Whichever miner completes the
challenge propagates the block to the rest of the network where each node
adds it to their local view of the blockchain. Nodes that submit transactions
but do not mine blocks, are called \emph{non-miners}. Note that while the purpose
of miners is to mine blocks, they may also submit and mine their own
transactions.

The ordering of transactions within a block is important. Each transaction is
crafted to operate on a specific blockchain state, but when a transaction executes, it changes
this state. Therefore, a transaction's execution can depend on the set of
transactions preceding it both in previous blocks and within the same block. Miners have full
discretion as to intra-block ordering. The default strategy is to sort pending
transactions in the mempool in descending order by fees-per-byte. This strategy
optimizes profit when transactions are considered individually.

A transaction may include a \emph{smart contract}~\cite{szabo1997}; an executable set of
instructions for the Ethereum Virtual Machine (EVM). A smart contract will often
specify the terms of a cryptocurrency transaction (e.g. source, destination, amount,
conditions, etc.), but may also specify arbitrarily complex programs.

As the EVM instruction set is Turing-complete~\cite{ethereum}, Ethereum
introduces the notion of \emph{gas} to guarantee program termination. Gas works
by charging a fee per executed instruction. The cost of an instruction
in \emph{gas} is constant, however, when a transaction is submitted to the
network, the submitter must choose an exchange rate of Ether (ETH) to gas.
This is called the \emph{gas price}. Transaction submitters are incentivized to choose
competitive gas prices so that miners select their transactions when mining blocks.
Miners run the smart contracts, and, upon successfully mining a block, receive,
as a fee, the gas used for each transaction within the block. If a contract runs
out of gas, the miner gets to keep the gas fees, but rolls back any side-effects
of the contract.


\subsection{DEXes, Frontrunning, and MEV}

Like a traditional market, cryptocurrency markets rely on exchanges. An exchange is
somewhere traders can swap assets, and usually offers some form of protection
for the involved traders. A popular type of exchange for cryptocurrencies is a
\emph{decentralized exchange} (DEX). In a DEX, the logic of the exchange is
relegated to a smart contract. In one type of DEX, called an \emph{Automated Market
  Maker} (AMM), the DEX contracts hold currency reserves themselves instead of
being maintained by a trusted bookkeeper.

For decades, investors in traditional finance markets have sought creative modes
of profit that give them a competitive advantage. One such method is called
\emph{frontrunning}~\cite{Bernhardt_Taub_2008}. Frontrunning is a predatory
investment strategy where an investor learns of an attempted trade in an
exchange by a third party. Then uses privileged access to the trading platform
to execute trades faster than their victim.

\subsubsection{Transaction Ordering}

In the context of Ethereum, a miner can engage in frontrunning by simply
choosing to place one of their own transactions ahead of the target victim
within the same block. A non-miner can, in fact, achieve the same effect by
increasing the gas price such that a rational miner will naturally order the
non-miner's transaction ahead of the victim. The inverse of this is
\emph{backrunning} in which the MEV extractor wishes their transaction to be ordered
after the victim. Backrunning can be accomplished by analogous methods for both
miners and non-miners.

\subsubsection{MEV Extraction}
In their seminal 2020 paper, \citet{daian2020flash} showed that frontrunning and
backrunning are common occurrences in Ethereum DEXes along with several other types of
predatory transaction behaviors. They collectively called these behaviors
\emph{Miner Extractable Value} (later changed to \emph{Maximal} Extractable
Value, and usually referred to as MEV).
The idea behind MEV extraction is for profit seekers to discover financial
instabilities in existing protocols and to craft transactions that exploit these
instabilities in order to extract monetary value. MEV extraction may leverage and
combine transaction ordering primitives.

MEV in the DeFi space is made possible by the incentives behind transaction
ordering within blocks. In this paper, we will discuss three MEV strategies, and
their impacts. The strategies are \emph{sandwiching}, \emph{arbitrage MEV}, and
\emph{liquidation MEV}\cameraAdd{~\cite{zhouHighFrequency2021,qinQuantifyingBlockchainExtractable2021}}.

\paragraph{Sandwiching} 
Sandwiching is a classic trading strategy and is well-known in the traditional
financial world. However, it can be also be leveraged for MEV extraction.
An MEV extractor starts by monitoring the mempool for pending transactions that
are about to trade large sums of a particular asset. A large transaction
will result in a fluctuation in the price of the asset. Knowing this, the
MEV extractor then crafts a so-called ``sandwich'', by surrounding this large
transaction with two of its own transactions.
In the first transaction, the MEV extractor frontruns the large transaction in
order to buy or sell some quantity of the asset before the price of the asset
fluctuates. In the second transaction, the MEV extractor backruns the large
transaction in order to either buy back the original asset at a lower price, or
sell the newly acquired asset for a higher price. In both cases the MEV
extractor makes a profit due to the price difference.

We model sandwiching as follows.
\begin{defn}
Consider a victim transaction
$V$ and sandwich transactions $t_{1}$ and $t_{2}$ originating from the
same address. Transactions $t_{1}$ and $V$ are
exchanges from currency $C_{x}$ to currency $C_{y}$, and $t_{2}$ is
an exchange from $C_{y}$ to $C_{x}$. All three transactions are
within the same block $B$, i.e., $t_{1}, t_{2}, V \in B$. A sandwich
MEV is said to have occurred if $t_{1} \prec V \prec t_{2}$.
\end{defn}

\paragraph{Arbitrage MEV}
Arbitrage is the process of trading assets simultaneously across different
exchanges in order to profit from price differences; it is typically
considered benign as it balances out price differences across
exchanges and keeps the market stable.

An MEV extractor can perform MEV on arbitrage transactions by either following a \emph{passive} or
\emph{proactive} strategy. When following a passive strategy, the MEV extractor
monitors the current blockchain state and compares prices for different assets
across multiple exchanges and only executes an arbitrage if the expected revenue
of buying an asset on one exchange and selling it on another
exceeds the expected transaction costs.

In the proactive strategy, the MEV extractor monitors the mempool seeking either
a pending arbitrage transaction or a large pending trade. When a pending
arbitrage transaction is detected, the MEV extractor simply copies the
transaction and pays higher transaction fees in order to frontrun the existing
transaction and claim the profits of the arbitrage transaction. When a large
pending trade is spotted, the MEV extractor will first check if the large trade
will result in a price difference across different exchanges and only then craft
an arbitrage transaction around the large trade by attempting to backrun the
large pending trade with its own arbitrage transaction.

We formally define proactive arbitrage MEV as follows.
\begin{defn}
Consider a victim arbitrage
transaction $V$ that is unpublished and has only been propagated in the public
mempool. $V$ is an exchange of a single currency $C$ between two exchanges
$E_{1}$ and $E_{2}$, we write this as $E_{1}(C) \to E_{2}(C)$. We denote the
price of $C$ on an exchange as $P(E_{n},C)$. Arbitrage can occur if
$P(E_{1},C) + F_{vic} < P(E_{2},C)$, where $F_{vic}$ is the mining fee paid by
the victim. If an MEV extractor learns of such an opportunity in the mempool, it
can submit the same transaction $E_{1}(C) \to E_{2}(C)$ with a higher fee (i.e.,
$F_{mev} > F_{vic}$) as long as $P(E_{1},C) + F_{mev} < P(E_{2},C)$ still holds.
\end{defn}

\paragraph{Liquidation MEV}
Lending and borrowing assets is one popular use case of DeFi.
Lenders aim to profit by charging interest on capital they lend to
\emph{lending pools} (e.g. Aave~\cite{aave} or Compound~\cite{compound}).
Borrowers borrow assets from these lending pools by agreeing to pay
interest on the loan and by proffering a security in the form of collateral
worth more than the borrowed asset.

Complicating matters, the value of a collateral can fluctuate over time.
If the price of the collateral drops by a certain critical value, it becomes
worth less than the value of the borrowed asset. In order to prevent the
collateral from reaching this critical value, lending pools allow anyone to
\emph{liquidate} loans. Liquidation is when a user \cameraAdd{(distinct from the
  borrower)} repays the debt \cameraAdd{on behalf of the borrower} in exchange
for receiving the collateral at a discounted price. \cameraAdd{The discount on
  the collateral incentivizes liquidation, so the lender can avoid losing money
  on the loan.} When the value of a
collateral approaches the critical value, the lending pool marks the loan as
``unhealthy'' and makes the loan available for liquidation.
After liquidation, the loan becomes ``healthy'' again, and the liquidator can
profit by selling the collateral in the market.

There exist two different types of liquidation mechanisms: \emph{fixed
spread-based} liquidations and \emph{auction-based} liquidations.
Fixed spread-based liquidations are settled within one blockchain transaction
and follow the first-come-first-served principle: whoever offers to liquidate a
loan first receives the collateral. On the other hand, auction-based
liquidations are initiated by interested liquidators who provide bids with the
highest bid receiving the loan's collateral. An auction may last several hours
and is non-atomic as it may require liquidators to interact with the lending
platform via multiple transactions.

Due to their atomicity, fixed spread-based liquidations are a prime target
for MEV extraction. Similar to arbitrage, a would-be MEV extractor can either
follow a \emph{passive} or \emph{proactive} strategy to perform liquidations.
When following a passive strategy, the MEV extractor only scans the current
blockchain state for liquidation opportunities and attempts to frontrun competing
liquidators. In the proactive strategy, the MEV extractor monitors the
mempool for either of two types of opportunities. The first are pending liquidation
transactions, which the extractor can then copy and frontrun; the second are
pending transactions that will create a liquidation opportunity when mined (\eg
an oracle price update that renders collateral open for liquidation), for which
the extractor then crafts a liquidation transaction and backruns the transaction
that created the liquidation opportunity.

We formally describe fixed spread-based, proactive liquidation as follows.
\begin{defn}
Consider a collateral $C$ which is leveraged against a loan with value $v_{L}$.
The current price of the collateral at time $t$ is $P(C, t)$. When the loan is
taken out, $P(C, t_{0}) = v_{L}$. If at some future time
$P(C, t_{n}) < v_{L}$, the loan is released by the lending pool for
liquidation.

An MEV extractor may attempt to frontrun a liquidation $tx_{1}$ by submitting
a transaction $tx'_{1}$ (a copy of $tx_{1}$) where $tx'_{1} \prec tx_{1}$ in the
next block. As in sandwiching, this is accomplished by setting
$tx'_{1}.fee > tx_{1}.fee$.
\end{defn}

\subsection{Flash Loans}

Flash loans are a concept unique to DeFi. They are loans taken and repaid within
the same transaction. Flash loans are possible because a single transaction can
provide guarantees on the collective execution of multiple smart contract
functions, \ie they can ensure that either all function calls succeed or the
transaction is reverted. The advantage of Flash loans is that they allow users
to borrow any assets available in a lending pool without requiring any
collateral. If the user is not able to pay back the
loan plus interest at the end of the execution of the transaction, then the
transaction is reverted, meaning any state changes that occurred during
execution are rolled back.

As flash loans are bound to the execution of a single
transaction, they cannot be leveraged by MEV extractors for sandwiching.
However, they can be leveraged by MEV extractors to perform arbitrage MEV and
liquidation MEV using large, borrowed assets. The MEV extractors are only required to own
enough assets to pay back the interest and execution costs.

\subsection{Private Transactions}

In Ethereum, most transactions are propagated when users submit their
transactions to Ethereum nodes which then gossip them to other nodes in the
network~\cite{kiffer2021gosssip}. Eventually, every node that is part of Ethereum's peer-to-peer
network will receive the transaction.

In Ethereum, transactions are not encrypted, so everyone can
inspect the contents of a transaction. This, combined with the fact that
transactions are publicly propagated, allows users to simply connect to an
Ethereum node and monitor which transactions are pending and analyze their
purpose. This openness, combined with the fact that miners predictably order
their transactions by gas price, ultimately enables frontrunning and MEV
extraction.

One way to mitigate these complexities is to
send transactions directly to a trusted miners who will not further propagate
the transaction and will mine it secretly. These secretly submitted transactions
are called \emph{private transactions}. Projects such as \fb or the Eden
Network~\cite{eden} try to optimize this process of creating private agreements
with miners and establishing so-called \emph{private pools}. Transactions within
these pools are only visible and forwarded to trusted miners.

There are typically three reasons why users might want to use private
transactions: miner payouts, transaction privacy, and frontrunning/backrunning.
Private transactions are useful for miner payouts in mining pools because their
pool does not need to include any transaction fees in its payout transaction. It
can get away with this because it knows that all miners in the mining pool have
a vested interest in the transaction being included.
Private transactions are also useful for maintaining some degree of transaction privacy.
Users may want to hide the intention of their transactions, for
example, when performing a trade they do not want to be affected by MEV
extraction, or when performing MEV extraction they do not want to be frontrun,
themselves, by competitors.
Finally, private transactions are useful for would-be frontrunners who wish to
maintain their competive advantage by hiding their practices from other
frontrunners.

\subsection{\fb}\label{sec:flashbots}

\fb is the name of an ecosystem of projects with the stated goal of limiting
frontrunning's negative externalities. It does this primarily through two
initiatives: \texttt{MEV-geth}~\cite{mev-geth},
and \texttt{MEV-inspect}~\cite{mev-inspect}.

\begin{figure}
    \centering
    \includegraphics[width=0.7\columnwidth]{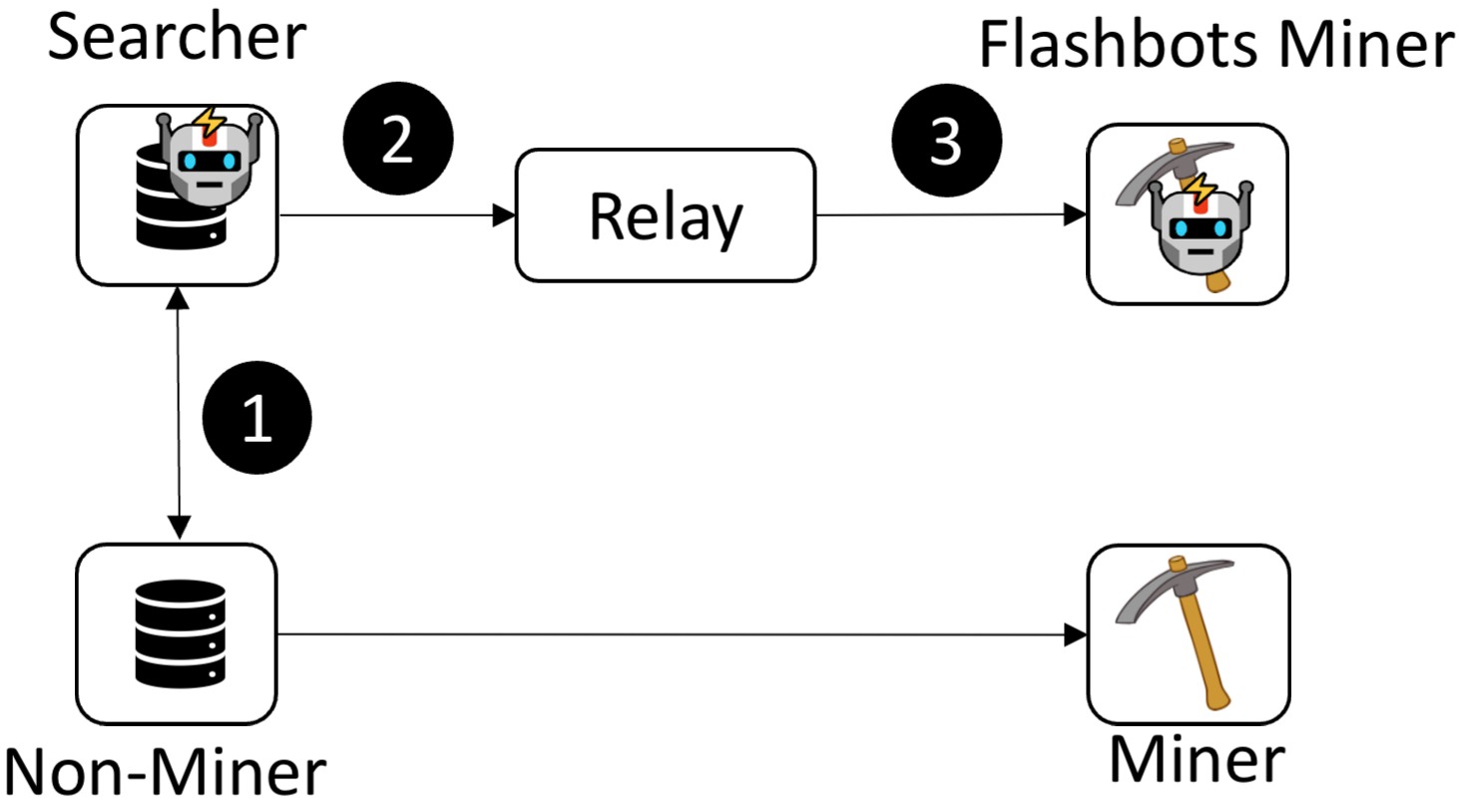}
    \caption{\fb architecture. Searchers, a subset of non-miners, send
      transaction bundles to relays who then forward them to participating \fb
      miners. Non-\fb nodes are unaware of these bundles.}
    \label{fig:flash_arch}
\end{figure}

The \fb ecosystem works by creating a private transaction pool, which can only
be accessed by nodes assuming one of three delineated roles. The roles are:
\emph{searchers}, \emph{relays}, and \emph{miners} (\Cref{fig:flash_arch}).
Searchers are non-miners who listen for public transactions propagated by peers
(Step~\step{1}). When they identify MEV opportunities, they build a \emph{bundle} of
transactions, which is an immutable, atomic set of transactions---either all
transactions in the bundle are executed \emph{in order}, or none of them are.
\cameraAdd{Bundles are a concept exclusive to \fb.} The
bundle includes a fee that is paid out to the miner\cameraAdd{, but the MEV profit from the submitting
searcher's transactions are kept by that searcher.}
There are three
different bundle types:
(1) \emph{miner payout}, which include transactions used by the miner to
pay users of a mining pool,
(2) \emph{rogue}, which include transactions introduced by the miner and not
broadcast even with \fb, and
(3) \emph{flashbots}, which follow the standard dataflow described below.

After the searcher has identified MEV opportunities, they then forward the
bundle to the relays (Step~\step{2}) instead of gossiping to the public mempool
as is the case with non-\fb transactions. \cameraAdd{\fb bundles (and their
  constituent transactions) thus
  remain visible only to \fb miners; they become visible to the rest of the
  public Ethereum network only after they have been mined into a block.} Relays collect bundles from searchers
and forward them to miners---they exist primarily as DoS protection for
unprepared miners. \fb plans to add the ability for any node to serve as a relay
in the future, but currently, there is only one relay in the system, run by the
\fb project itself.

Miners collect bundles from relays (Step~\step{3}), and mine whichever
bundles are most profitable for them\cameraAdd{---only a single bundle can be included per
block to guarantee the searcher's intent is satisfied. Since bundles are just an
ordered set of transactions, they do not qualitatively differ from any other
Ethereum transactions after mining.} Miners can participate by running
the {\fb}-provided \texttt{MEV-geth}---a fork of the reference go-ethereum
(\texttt{geth}) client~\cite{go-ethereum}.
Miners can choose which bundles to mine, most likely based on potential for
profit, and they can choose not to mine bundles at all. \cameraAdd{All choices
  for miners are devoid of pecuniary risk as MEV is a pure profit opportunity
  for them.} However, if they choose
to mine a bundle, they cannot in any way modify that bundle. Equivocating on a
bundle leads to a permanent ban from using the \fb infrastructure.

\fb is able to threaten banning, because they control access to the system. Any
miner or searcher wishing to participate in \fb must first apply through the \fb
web portal. They supposedly permit anyone to join who agrees to honor the above
invariants, however this has not been empirically evaluated. \cameraAdd{Upon
  passing a review by the \fb project, a miner is authorized to receive
  bundles from the single operational relay. Abuse of the system rules
  will result in revocation of a miner's authorization.}

\subsubsection{Goals of \fb}
\fb was specifically designed with three goals in mind. All design decisions are
allegedly in service of these three goals. We describe them here in no
particular order.

\begin{goal}[Illuminate the Dark Forest] \label{goal:illuminate}
    \fb claims that the MEV problem is difficult to measure and quantify. As such,
    they are spearheading a number of measurement
    initiatives~\cite{transparency,mev-explore}.
    The end goal: more transparency into how much MEV is occurring, and to what
    degree.
\end{goal}

\begin{goal}[Democratize MEV Extraction] \label{goal:democ_mev_extract}
    Form a neutral, public, open-source, and permissionless
    infrastructure for MEV extraction, with the intent that MEV extraction can be
    done by any node in Ethereum---not just those that can afford expensive
    infrastructure.
\end{goal}

\begin{goal}[Distribute Benefits] \label{goal:dist_bene}
    MEV primarily benefits traders and miners, but involves most participants in
    the Ethereum ecosystem. This goal of \fb is to help all participants in
    Ethereum profit from MEV.
\end{goal}

\section{Data Collection and Processing}\label{sec:datasets}
To evaluate the efficacy of \fb as a solution to the MEV crisis, we adopt a
data-driven methodology. We describe our data collection in this section as well
as our methods for extracting insights from these data. In the service of open
science, we make our datasets and collection code openly
available\footnote{\url{https://github.com/a-flashbot-in-the-pan}}.

\begin{figure}
    \centering
    \includegraphics[width=\columnwidth]{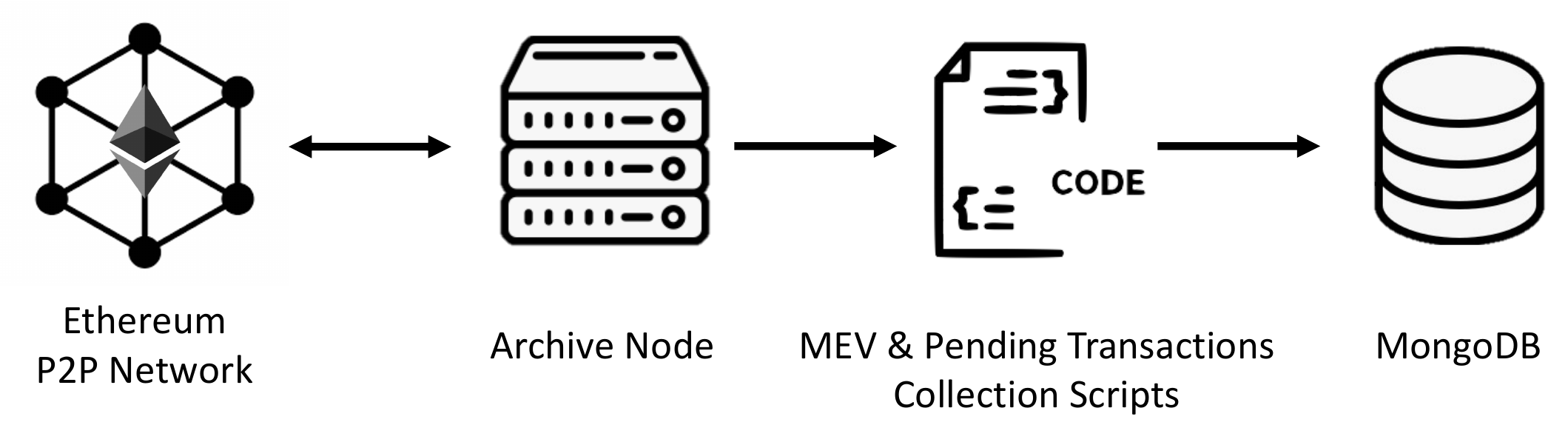}
    \caption{Measurement architecture.}
    \label{fig:architecture}
\end{figure}

\Cref{fig:architecture} provides an overview of our measurement setup for
collecting MEV-related data as well as pending transactions on the network. Some
of our data was collected using an Ethereum \emph{archive node}, which we setup
using \texttt{go-ethereum} \cite{go-ethereum}. An archive node provides a
complete history of all state changes on the Ethereum blockchain, which allowed
us to query data on any published block.

Our data collection focused on the block range between \num{10000000} and
\num{14444725} (May 4th, 2020 to March 23rd, 2022). All data collection used
a machine with \SI{18}{TB} SSD storage, \SI{128}{GB} memory and \num{10} Intel
(R) Xeon (TM) L5640 CPUs with \num{12} cores each and clocked at \SI{2.26}{GHz},
running 64-bit Ubuntu 16.04.6 LTS. We store our collected data inside a MongoDB
database.

\subsection{MEV}

\begin{table}
  \centering
  \begin{adjustbox}{width=\columnwidth,center}
  \begin{tabular}{l r r r r}
\toprule
\textbf{{MEV Strategy}} &  \textbf{Extractions} & \textbf{Via \fb} & \textbf{Via Flash Loans} & \textbf{Via Both} \\
\midrule
Sandwiching  & 1,020,044  &  485,680  (47.61\%) & 0 (0\%)  & 0 (0\%)  \\
Arbitrage  &  3,462,678  & 916,709 (26.47\%)  & 10.155 (0.29\%) & 1.089 (0.03\%) \\
Liquidation  & 32,819 & 9,191 (28.01\%) & 1.672 (5.09\%) & 132 (0.4\%) \\
\midrule
\textbf{Total}  & \textbf{4.515.541} & \textbf{1.411.580 (31.26\%)}  & \textbf{11.827 (0.26\%)} & \textbf{1.221 (0.03\%)} \\
\bottomrule
  \end{tabular}
  \end{adjustbox}
  \caption{MEV dataset overview.}
  \label{tbl:mev_extraction}
\end{table}

%
%

We developed scripts that leverage heuristics presented by previous
works~\cite{ferreira2021frontrunner,qinQuantifyingBlockchainExtractable2021} to
measure extracted MEV. We also leveraged techniques described by
\citet{wang2020flashloans} to detect flash loans and use public \fb data to
identify if the MEV has been extracted via \fb.

\subsubsection{Sandwiches}\label{sec:sand_detect}

We measured sandwiches by crawling token transfer events from our archive node
and applying the heuristics developed by \citet{ferreira2021frontrunner} for
detecting insertion frontrunning (\ie sandwiching). 
\cameraAdd{The heuristics by \citet{ferreira2021frontrunner} assume that the attacker buys and sells the same type of tokens as the victim across two different transactions where the amount purchased and sold by the attacker is almost identical and where the gas price of the attacker's first transaction is higher than the victim's transaction.}
Our script is capable of
detecting sandwiches across the following exchanges: Bancor~\cite{bancor}, SushiSwap~\cite{sushiswap}, and Uniswap V1, V2 and V3~\cite{uniswap}.

We computed the profit that the MEV extractor made by deducting the costs
of the sandwich transactions from the gain made through
those sandwich transactions. The costs were computed as the sum of the transaction fees for
both sandwich transactions and any tips that the MEV
extractor payed to the miner via coinbase transfers (these tips only apply when
the MEV extractor used \fb).

The gain was computed as the difference between
the ether used to purchase the tokens and the ether obtained
from selling the tokens. We collected \num{1020044} sandwich attacks, where \num{485680}
(\SI{47.61}{\percent}) of these sandwich attacks were performed by using \fb (see
\Cref{tbl:mev_extraction}).

\subsubsection{Arbitrage}

We measured arbitrage MEVs by crawling token swap events from our archive node
and applying the heuristics developed by
\citet{qinQuantifyingBlockchainExtractable2021}. \cameraAdd{The heuristics by
  \citet{qinQuantifyingBlockchainExtractable2021} assume that an arbitrage must
  include more than one swap event and that all swap events must be included in
  a single transaction thereby forming a closed loop.} Our script detected
arbitrages across the following exchanges: 0x Protocol~\cite{0x}, Balancer~\cite{balancer}, Bancor~\cite{bancor}, Curve~\cite{curve}, SushiSwap~\cite{sushiswap}, and Uniswap V2 and V3~\cite{uniswap}. We compute the profit that the MEV
extractor made, by deducting the costs of the arbitrage transaction from the gain
made through the arbitrage transaction.

The costs were computed as the sum of the transaction fees for the arbitrage MEV
transaction and any tips that the MEV extractor paid to the miner via coinbase
transfers (these tips only apply when the MEV extractor used \fb).

The gain was computed as the sum of the assets that were left after the swaps
were executed. Assets may include ether or tokens. To keep our analysis
consistent, we convert the value of the tokens to ether. To do so, we leverage
CoinGecko's API~\cite{coingecko}.
We collected \num{3462678} arbitrages, where \num{916709}
(\SI{26.47}{\percent}) of these arbitrages were performed by using \fb, \num{10155}
(\SI{0.29}{\percent}) by using flash loans, and \num{1089} (\SI{0.03}{\percent})
by using both \fb and flash loans (see \tablename~\ref{tbl:mev_extraction}).

\subsubsection{Liquidation}

We measured liquidation MEVs by crawling liquidation events from our archive
node for different lending platforms and extracting information from these events
such as liquidated debt and received collateral. Our script is
capable of detecting liquidations across the following lending platforms: Aave
V1 and V2~\cite{aave}, and Compound~\cite{compound}. \cameraAdd{We did not rely
  on previous works to detect liquidations. Concretely, our script searches for
  Aave's \texttt{LiquidationCall} event and Compounds's \texttt{LiquidateBorrow}
  event, which directly correspond to liquidations.}

We computed the MEV extractor's profit by
deducting the costs of the liquidation MEV transaction from the gain made through
the liquidation transaction.
Like in the arbitrage case, the costs were computed as the sum of the transaction fees of
the liquidation MEV transaction, the value of the liquidated debt, and any tips
that the MEV extractor payed to the miner via coinbase transfers (the last one only
applies if the MEV extractor performed the liquidation through \fb.

The gain was computed as the value of the received collateral. Similar to
arbitrage, we converted all collateral into its equivalent amount in ether using
the CoinGecko's API. We collected \num{32819} liquidations, where \num{9191}
(\SI{28.01}{\percent}) of these liquidations were performed by using \fb, \num{1672}
(\SI{5.09}{\percent}) by using flash loans, and \num{132} (\SI{0.4}{\percent})
by using both \fb and flash loans (see \tablename~\ref{tbl:mev_extraction}).

\subsubsection{Limitations}
\cameraAdd{Our measurements leverage heuristic-based MEV detection methods
  initially devised in prior research efforts~\cite{ferreira2021frontrunner,qinQuantifyingBlockchainExtractable2021}.
  Our results, therefore, suffer from the same limitations as those
  works, and should be considered a lower bound on the number of MEV instances.

  One such limitation in
  our methodology is that our sandwich MEV detection mechanism
  assumes that both sandwich transactions always occur within the same block.
  While this assumption is a boon to our ability to process the massive
  blockchain history efficiently---it is not strictly true. A single sandwich's
  transactions can be located on different blocks and still be successful (i.e.,
  profitable for the submitter). Our methodology will not detect this.

  Another limitation in our methodology is that we do not account for any other
  types of MEV. In this study we solely
  focus on the most well known and popular types: sandwiching, arbitrage, and
  liquidation. If other types exist, they would require their own unique
  detection methods and analyses.

  Finally, we performed our measurements on
  popular exchanges and lending platforms. However, other exchanges and platforms
  may also suffer from MEV extraction and were not covered in this
  study simply because they are less popular.}

\subsection{Pending Transactions}
  
We collected pending transaction via our archive node for a period of five
months, between November 8th, 2021 and April 9th, 2022. We developed a script
that used the standard Web3
API's~\cite{web3js} \texttt{web3.eth.subscribe}
function to subscribe to ``pendingTransactions'' events, which are triggered
upon receipt of incoming pending transactions, i.e, when new transactions
enter the mempool. Next, our script retrieved the full corpus of details on a
transaction via the Web3 API's \ttt{web3.eth.getTransaction} function and stored this
information inside a MongoDB database. We collected a total of \num{125660856}
public pending transactions.

\subsection{\fb\ \cameraAdd{API}}

In line with their transparency initiative, \fb provides a public
API~\cite{fb_block_api}, where users can download and
inspect all bundles that have been mined by miners participating in the \fb
network. The data is publicly accessible and contains information such as block
number, miner address, miner reward, and transactions. Each transaction is
labeled with a bundle ID and a bundle type.

We downloaded the entire list of \fb blocks until block \num{14444725}
and obtained a list of \num{1196218} blocks that were mined as \fb blocks. We
then used the transactions included in those blocks to identify and mark transactions
as \fb transactions in our analysis.

\subsection{Flash Loan \cameraAdd{Extraction}}

We identify flash loans by applying the techniques described by
\citet{wang2020flashloans}. These techniques rely on crawling our archive node
for specific events that are only triggered by lending platforms when a flash
loan has been successfully executed. Our script is capable of detecting flash
loans for the lending platforms Aave~\cite{aave} and dYdX~\cite{dydx}.

\section{\fb Usage}\label{sec:usage}


The \fb project went live in January 2021, with the first \fb block being mined
a few weeks later on February 11th, 2021 at block height \num{11834049}.
In this section, we analyze usage patterns since \fb' advent. In particular, we
focus our analysis on four attributes of usage:
(1) the number of \fb bundles,
(2) a comparison of \fb usage to non-\fb usage,
(3) participation and hashing power of \fb miners, and
(4) usage towards MEV transactions.

\subsection{\fb Bundles}



The core unit of \fb usage is the \emph{bundle}. According to the \fb protocol,
all \fb transactions must end up on the blockchain wrapped in such a bundle. To
understand how \fb is being used, we needed to analyze these bundles, which are
published on the blockchain and consequently made available through \fb' public
API.

In usage data collected from the public \fb API, we found a total of \num{3249003}
bundles included in \num{1196218} blocks. The minimum number of bundles observed
per \fb block is one and the maximum is 42. On average a \fb block contains
\num{2.71} bundles, with a median of two bundles per block. Bundles contain on
average \num{2.15} transactions, with a median of one. However, the largest bundle
that we observed contained \num{700} transactions---this bundle was included in
block \num{12481590}. After further inspection, we found that the 700
transactions were miner payouts from F2Pool (one of the largest mining pools in
Ethereum). \num{1993775} bundles or \SI{61.37}{\percent} of the bundles only
contain one transaction. These numbers suggest that there might not be many bundles
available to mine, or we would expect to see more blocks with more bundles.
The small number of bundles may indicate that either there are not many
searchers submitting bundles, or their are not many MEV opportunities. We expand
on these points in \Cref{sec:mev_usage}.



Not all bundle types (\Cref{sec:flashbots}) are equally common.
Only \SI{1.9}{\percent} or \num{61500} bundles were from \emph{miner payouts},
and only \SI{7.6}{\percent} or \num{247523} bundles were marked as \emph{rogue}.
The rest of the bundles, \SI{90.5}{\percent} or \num{2939980} were
\emph{flashbots} bundles; these include sandwiches, arbitrage MEVs,
liquidation MEVs, and other order-dependent trades. These measurements suggest
that the vast majority of \fb transactions are either submitted to extract MEV
or to protect against MEV extractors.

\subsection{\fb Blocks Vs. Non-\fb Blocks}

\begin{figure}
    \centering
    \includegraphics[width=\columnwidth]{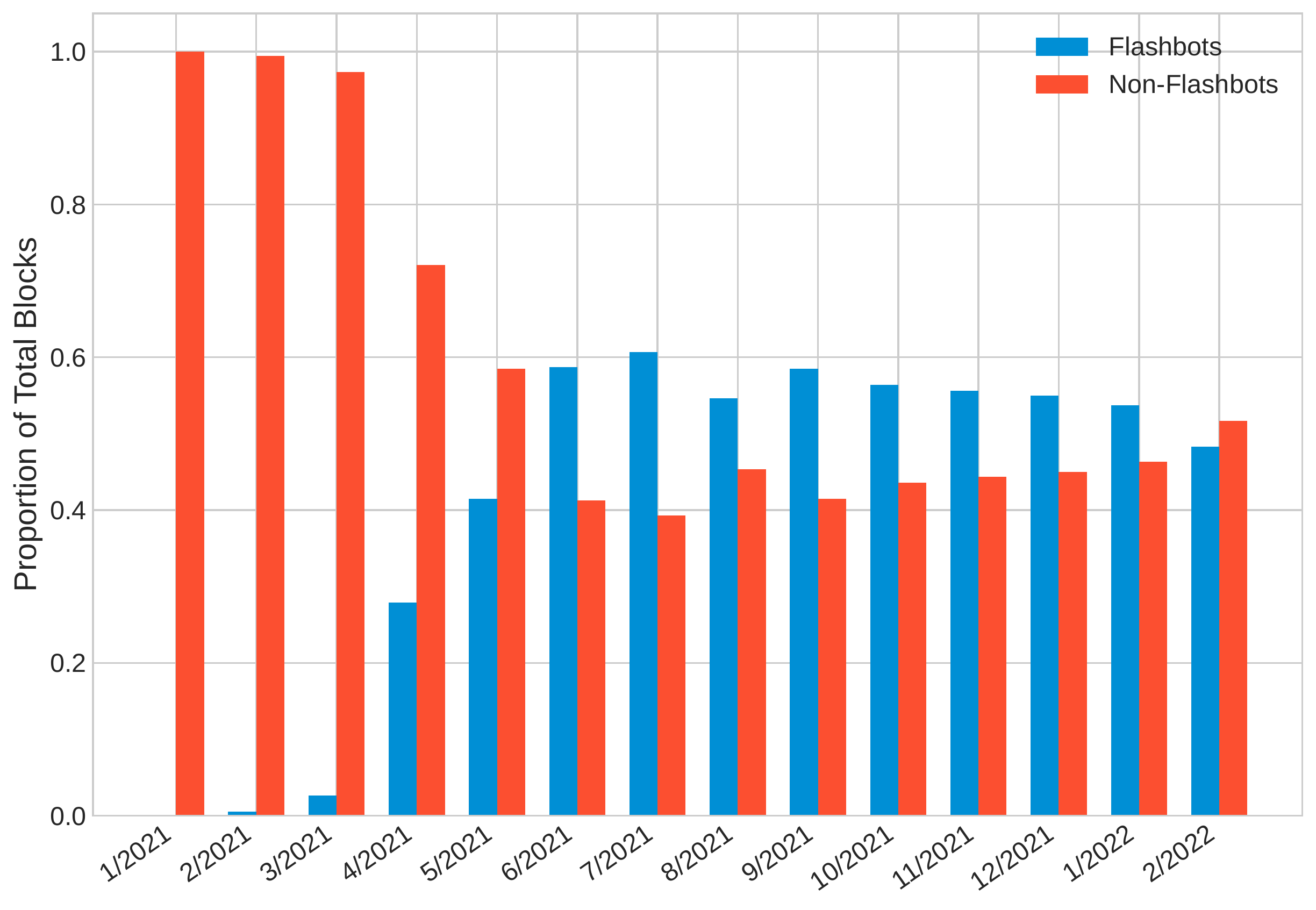}
    \caption{Proportion of \fb blocks mined compared to \emph{all} Ethereum
      blocks.}
    \label{fig:fb_hashpower}
\end{figure}

One metric of {\fb}' popularity is the ratio of Ethereum blocks that include
\fb transactions. This metric offers a high-level summary of \fb usage without
factoring in the behaviors of individual participants. We calculate these ratios
by summing the number of confirmed \fb blocks in each month and dividing by the
total number of Ethereum blocks in that same month. We show these ratios in
\Cref{fig:fb_hashpower}. We see that adoption started low, as expected, but rose
rapidly. The \fb ratio peaked in July 2021 at \SI{60.6}{\percent}. It then
hovered slightly above \SI{50}{\percent} for several months. As of February 2022
the ratio dipped back below non-\fb blocks (\SI{48.2}{\percent}).
As miners can be assumed to be rational, this decline indicates that fewer
bundles are likely available.


This calculation is complicated by the fact that
a \fb miner can decide arbitrarily to not include any \fb transaction.
However, we do not believe this is likely to happen.
The only two reasons that a \fb miner would not include any \fb transactions is
because either none are available, or because they are ignoring them (or copying
them for themselves). A rational miner will not ignore profitable transactions, and, we
argue, miners will not copy searcher transactions, because the searchers make
very little anyway (\Cref{sec:fb_profits}), and the miner would not want to risk
getting kicked out of \fb, thus depriving them of future profit.

\subsection{Hashing Power}

The hashing power of \fb miners is a strong indicator of \fb' popularity among
miners weighted by the contribution of those miners to the blockchain.
Measuring the hashing power of miners directly is not possible without access to
their hardware. We can, however, estimate a miner's hashing power (relative to
the rest of the Ethereum network) by counting the number of blocks mined
\emph{by that miner} over a controlled time span and dividing by the total
number of blocks in that same time span. This method yields the ratio of a
miner's hashing power compared to the rest of the network.

As we only care about \fb miners, we do this analysis \emph{only} for miners who have
mined a \fb block in each time span examined (one month). We assume that if a miner is
part of the \fb ecosystem (i.e., has mined at least one \fb block) then they
contribute to the total \fb hashing power in that span, even if some of their blocks
do not contain bundles. This is because a participating miner may not always
mine \fb blocks, even if they are still be actively considering available
bundles---bundles may not always be available, and some available bundles may not be
sufficiently profitable to justify inclusion.

\begin{figure}
    \centering
    \includegraphics[width=\columnwidth]{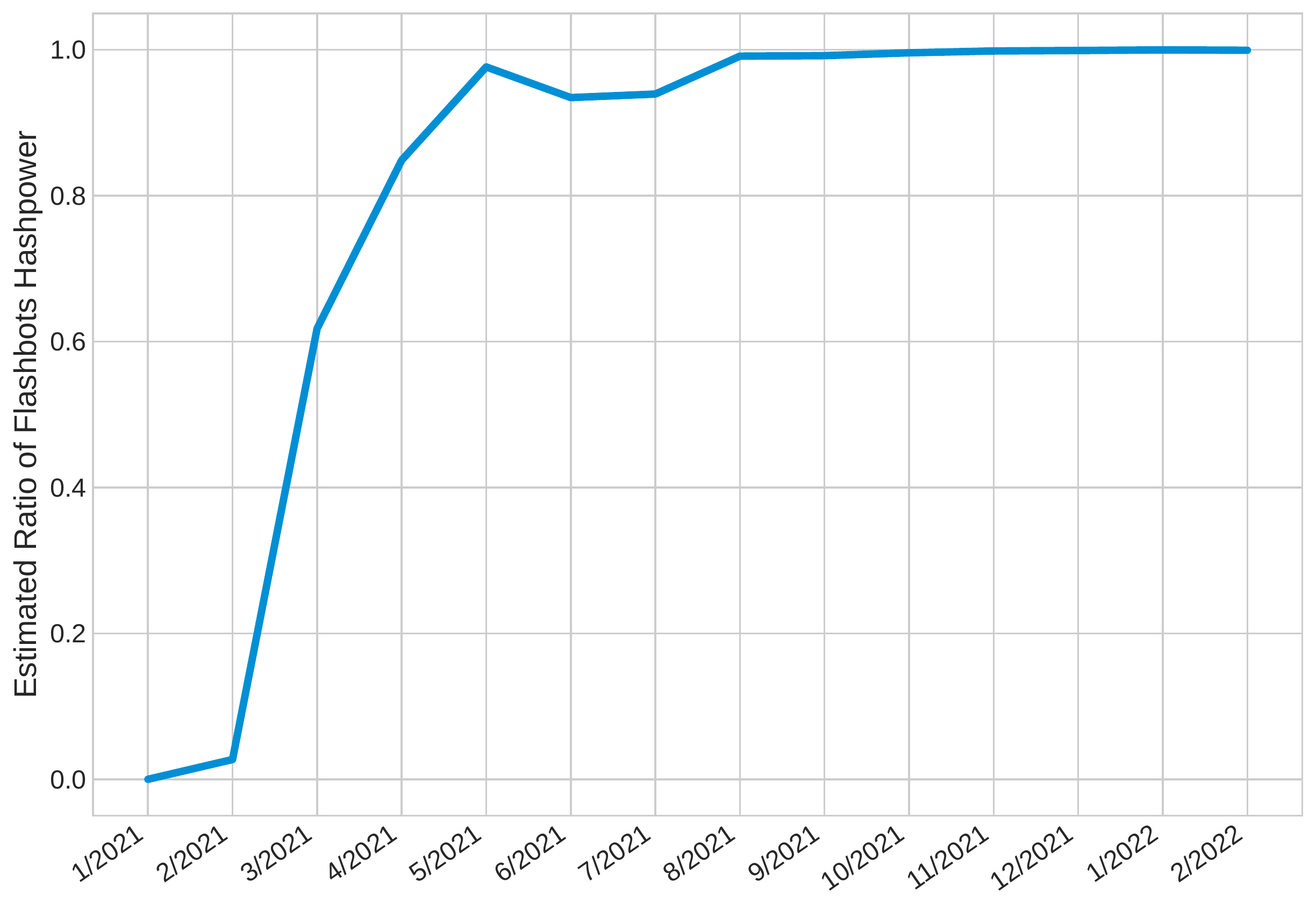}
    \caption{Estimated hashrate of \fb miners as a fraction of total Ethereum
      hashpower.}
    \label{fig:fb_hashpower_ratio}
\end{figure}

We present our findings in
\Cref{fig:fb_hashpower_ratio}. As expected, there were no \fb miners when the
system went live in January 2021. Miners joined in quickly however, and by
March accounted for \SI{61.7}{\percent} of the Ethereum hashrate. By May the
hash rate reached \SI{97.6}{\percent}. As of February 2022, it was hovering
around \SI{99.9}{\percent}\footnote{The \fb Transparency
Dashboard~\cite{transparency} estimates a hashrate
of \SI{74.5}{\percent}, however, we suspect this to be an underestimate. They
have not made their methodology for this measurement available.}. This means
that effectively \emph{all} prominent miners in Ethereum are enrolled in \fb.
This has serious implications for the future of Ethereum as it has become a
\emph{de facto} part of the Ethereum protocol. \cameraAdd{Though initially
  jarring, the result should not be surprising. With the huge profit potential
  for miners at \emph{no risk} (\Cref{sec:flashbots}) and requiring only a small software change (in
  many cases, a 1-for-1 replacement if using \ttt{MEV-geth}), adoption is a
  rational decision. Also note that the network's hashing power is heavily skewed~\cite{gencer2018decentralization},
  so it is likely that far fewer than \SI{99.9}{\percent} of miners have adopted
  \fb, but those who have are the ones with the most hashing power.}

\subsection{Participating Miners}

\begin{figure}
    \centering
    \includegraphics[width=\columnwidth]{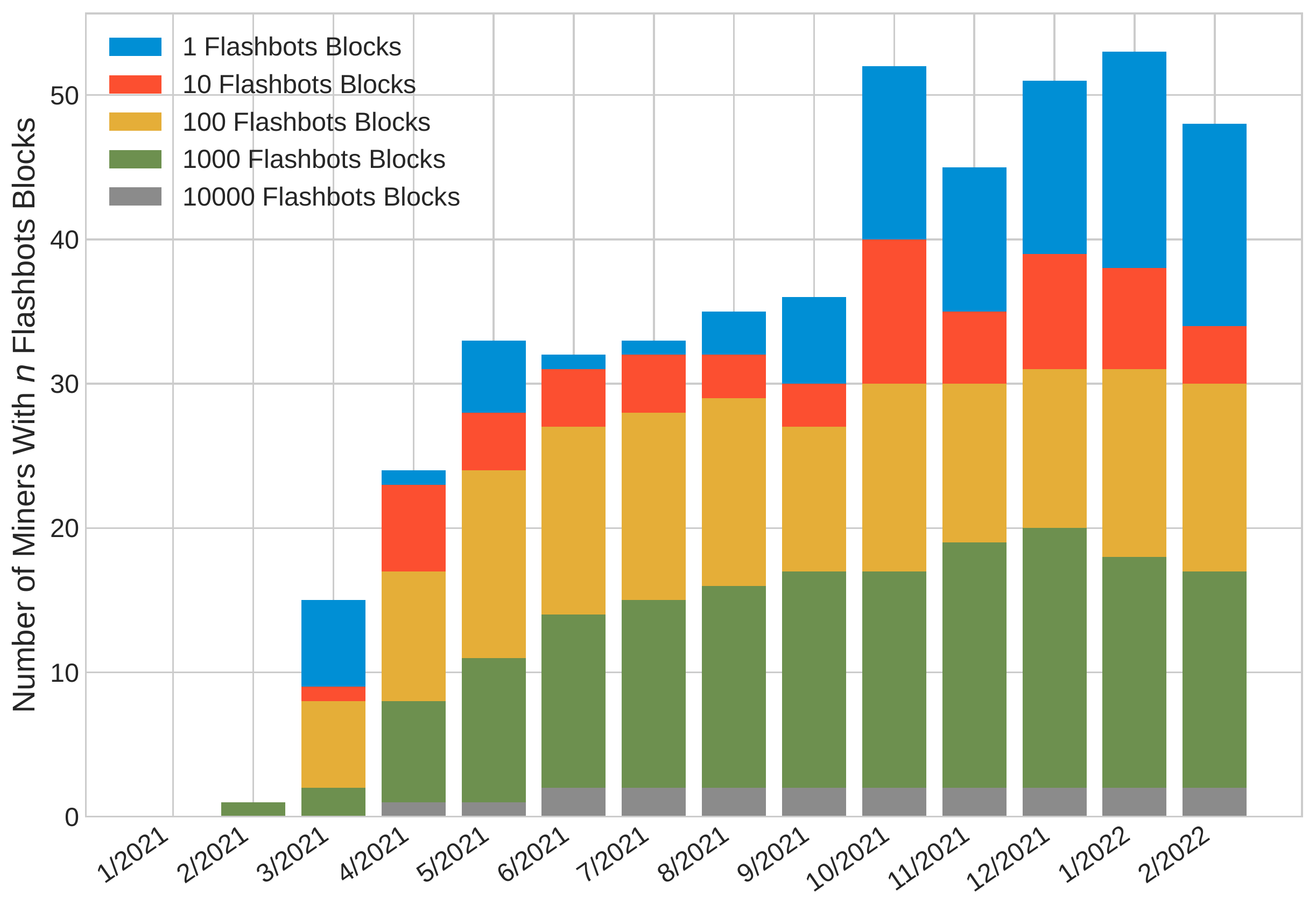}
    \caption{The number of miners who have mined $n$ blocks in each month of our
      analysis.}
    \label{fig:fb_blocks_per_miner}
\end{figure}

We next discuss the number of miners participating in \fb and how their hashing
power compares to each other. While the total hashing power of the network is
useful, it is also valuable to know how much individual miners are
contributing---this is especially true given \fb' calls for democratization
(\Cref{goal:democ_mev_extract}). We count the number of miners who have mined at
least $n$ blocks in each time span, where $n$ ranges logarithmically from
$10^{0}$ to $10^{4}$. We present these results in \Cref{fig:fb_blocks_per_miner}.
This figure shows that the distribution of blocks mined (and therefore, hashing
power) is long-tailed. This is consistent with previous
findings~\cite{gencer2018decentralization}. One or two miners mined over
\num{10000} blocks (depending on the month), but at \fb's peak adoption, this is
less than \SI{2}{\percent} of the total number of \fb miners. Additionally, no
month saw \fb blocks from more than \num{55} miners. In terms of \fb's goals,
this result does not seem to indicate a democratization of MEV
(\Cref{goal:democ_mev_extract}).

\subsection{MEV Usage}\label{sec:mev_usage}
\begin{figure}
    \centering
    \includegraphics[width=\columnwidth]{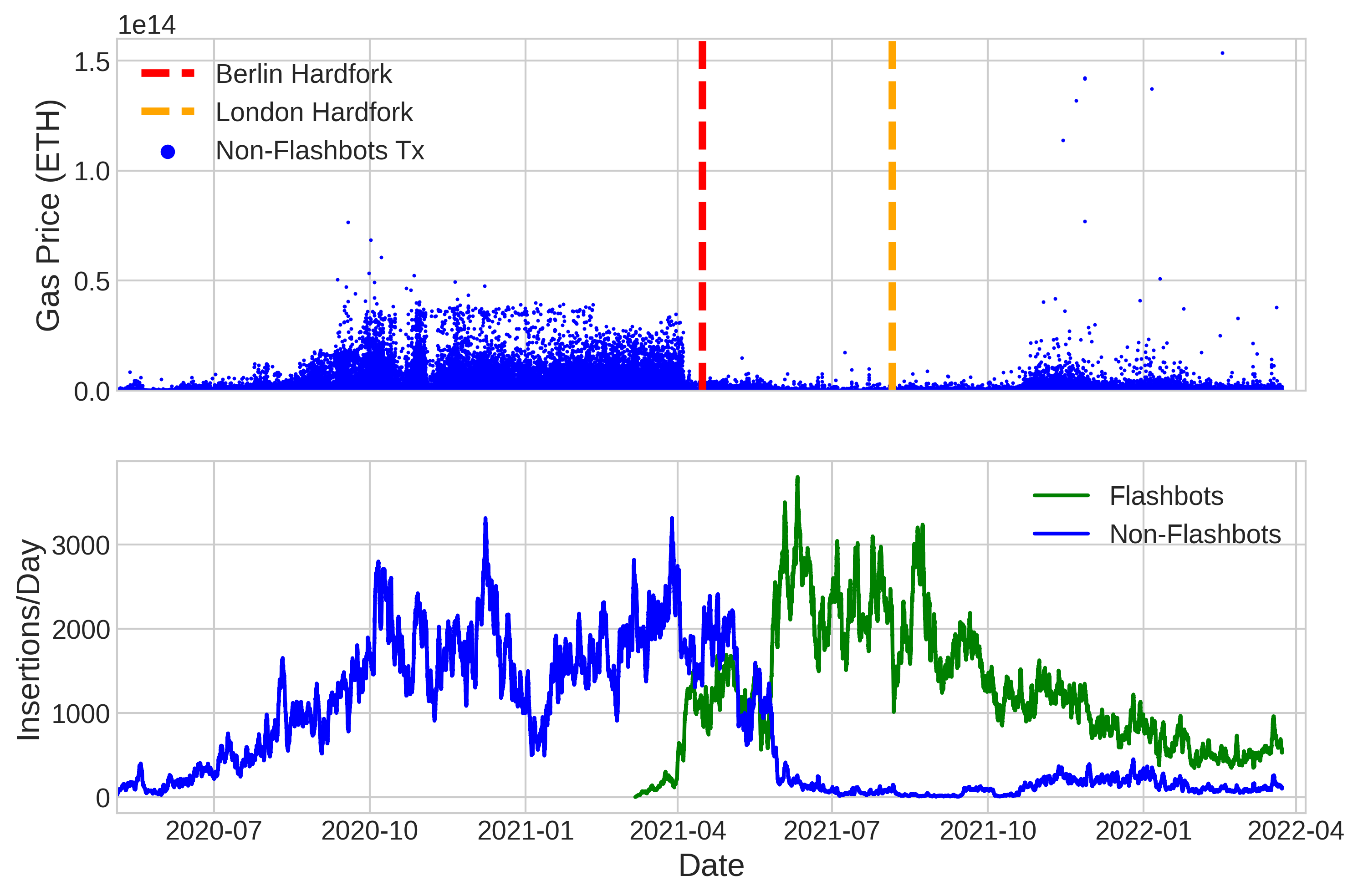}
    \caption{Correlation between sandwich transactions per day and gas price\cameraDel{,
      both y-axes are logarithmic}.}
    \label{fig:gas_price_txs_per_day}
\end{figure}

Here we measure number of sandwiches for both \fb and
non-\fb transactions over a 20-month period as well as the gas prices of these
transactions. We present our findings in \Cref{fig:gas_price_txs_per_day}. In
the top plot, we notice an interesting anomaly. There is a steep and sudden drop
in gas price in early April 2021. An astute reader might suspect this was caused
by one of Ethereum's hard forks. We have plotted the time of the two nearest hard forks
(code-named Berlin and London), but it is clear that they occurred well after
the steep decline and well before the uptick seven months later. However, when
plotted alongside the number of sandwiches (bottom plot in
\Cref{fig:gas_price_txs_per_day}), both \fb and non, we see that the precipitous
drop \cameraDel{coincides exactly}\cameraAdd{ correlates closely with the decline in
  \fb usage.}

Interestingly, we notice that both \fb and non-\fb transactions reduce in
frequency in September 2021. It is likely that this is due to the realization
that \fb is not profitable for non-miners (\Cref{sec:fb_profits}) along with the
concomitant rise in other private pools (\Cref{sec:private}).
We hypothesize that the steep valley in \Cref{fig:gas_price_txs_per_day} may coincide with the
popularity of \fb over time. \cameraAdd{The uptick in gas prices
  starting in September 2021 may be a result of decreased
  activity in \fb and simply be a return to pre-\fb behavior for non-miners.}

\subsubsection{MEV Types}
\begin{figure}%
    \centering%
    \begin{subfigure}[t]{\columnwidth}%
      \includegraphics[width=\linewidth]{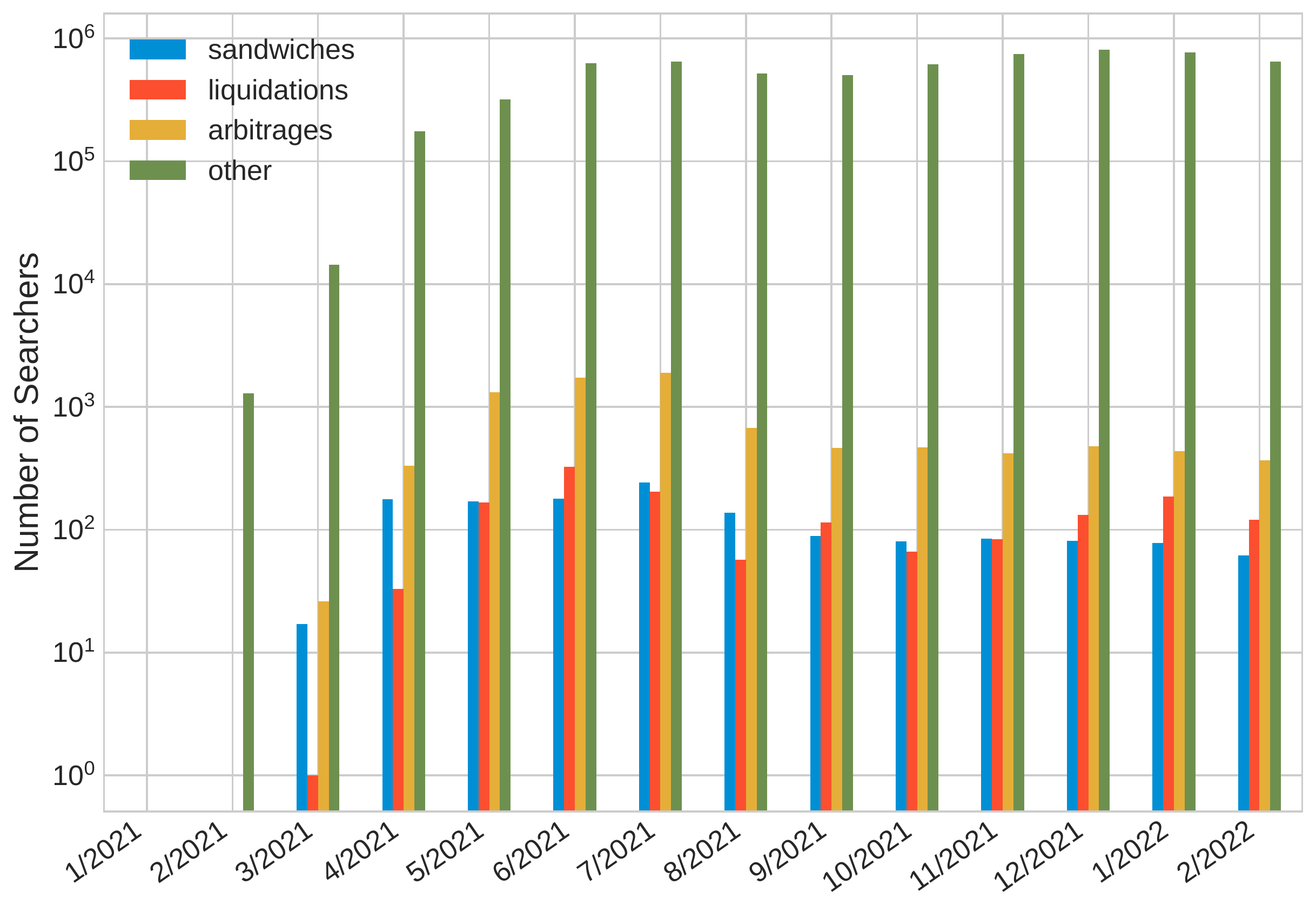}%
      \caption{Searchers by MEV type.}%
      \label{fig:searchers_by_type}%
    \end{subfigure}%

    \begin{subfigure}[t]{\columnwidth}%
      \includegraphics[width=\linewidth]{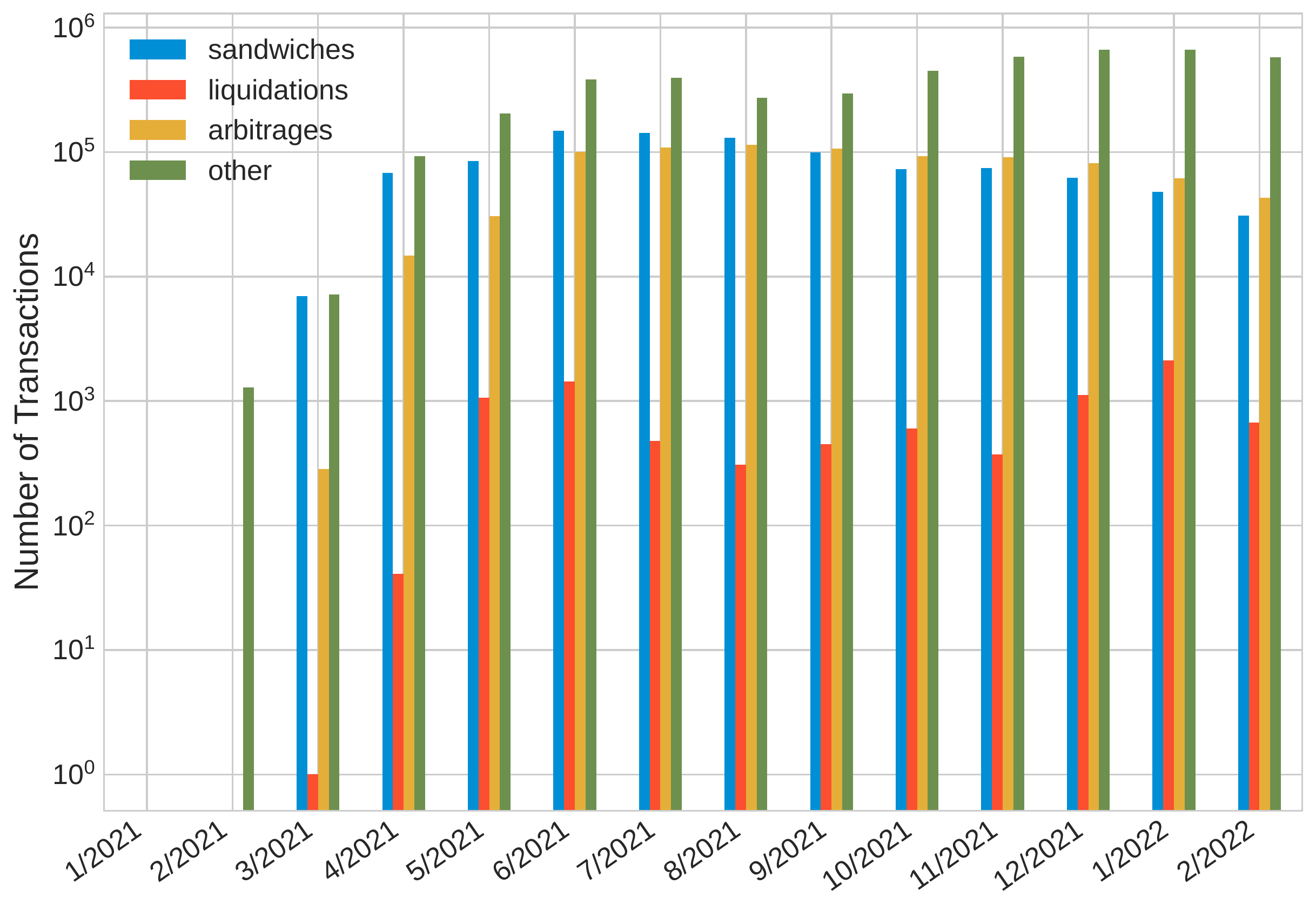}%
      \caption{MEVs by type.}%
      \label{fig:mevs_by_type}%
    \end{subfigure}%
    \caption{A breakdown of MEV transactions and searchers over town, separated
      by type\cameraAdd{, both y-axes are logarithmic.}}%
    \label{fig:mev_types}%
\end{figure}

Different types of MEV extractions will occur with different frequencies. In
\Cref{fig:mev_types} we track these frequencies over time. In the figure, we
present the three types of MEV extractions as well as an \emph{other} type, to
include \fb transactions that are not MEVs---these are likely either
order-dependent transactions or transactions that users wish to block MEV
extractors from profiting off of. In \Cref{fig:searchers_by_type}, we show the number
of \fb searchers that are engaging in each type of MEV. In all months, more
searchers include \emph{other} transactions than any of the MEV transactions (at
least two orders of magnitude more). We also see an interesting pattern within
each MEV type where they gradually increase through August 2021, before
decreasing and leveling out. This suggests that after some initial buzz, many
users left \fb for more profitable opportunities.

In \Cref{fig:mevs_by_type}, we see again that \emph{other} transactions are the
most popular type. However, the number of transactions of each type tend to be
more consistent than the number of searchers. This indicates that the
distribution may be heavily skewed with most MEV transactions are coming from
small number of searchers. We also note that the number of sandwiches and
arbitrage MEVs track each other closely, but there are far fewer liquidation
transactions. This is likely because there are simply fewer liquidation
opportunities.

\section{\fb Goals}\label{sec:audit}

The \fb project has established lofty goals (Benefit Distribution,
Democratization, and Illumination). In this section, we evaluate how effectively
it has achieved Benefit Distribution and Democratization. We discuss
Illumination in \Cref{sec:conclusion}.

\subsection{Profit Distribution}\label{sec:fb_profits}

\Cref{goal:dist_bene} of \fb is to distribute the benefits of MEV extraction. As
such, \fb aims to give non-miners the opportunity to profit off of MEV.
This is a reasonable proposition as MEV extracts value from \emph{all} participants in the
Ethereum ecosystem, so it makes sense to try to distribute profits among
those members of the community. Concretely, this implies that
miners, which are already flush with resources (e.g. money, powerful servers,
etc.) should make a lower share of the profits as compared to non-miner
participants (i.e., \emph{searchers})\footnote{This should not disincentivize miners,
per se, as long as they make a higher cumulative profit.}. However, our
measurements show that this is emphatically not the case.

\begin{figure}%
    \centering%
    \begin{subfigure}[t]{\columnwidth}%
      \centering
      \includegraphics[width=0.8\columnwidth]{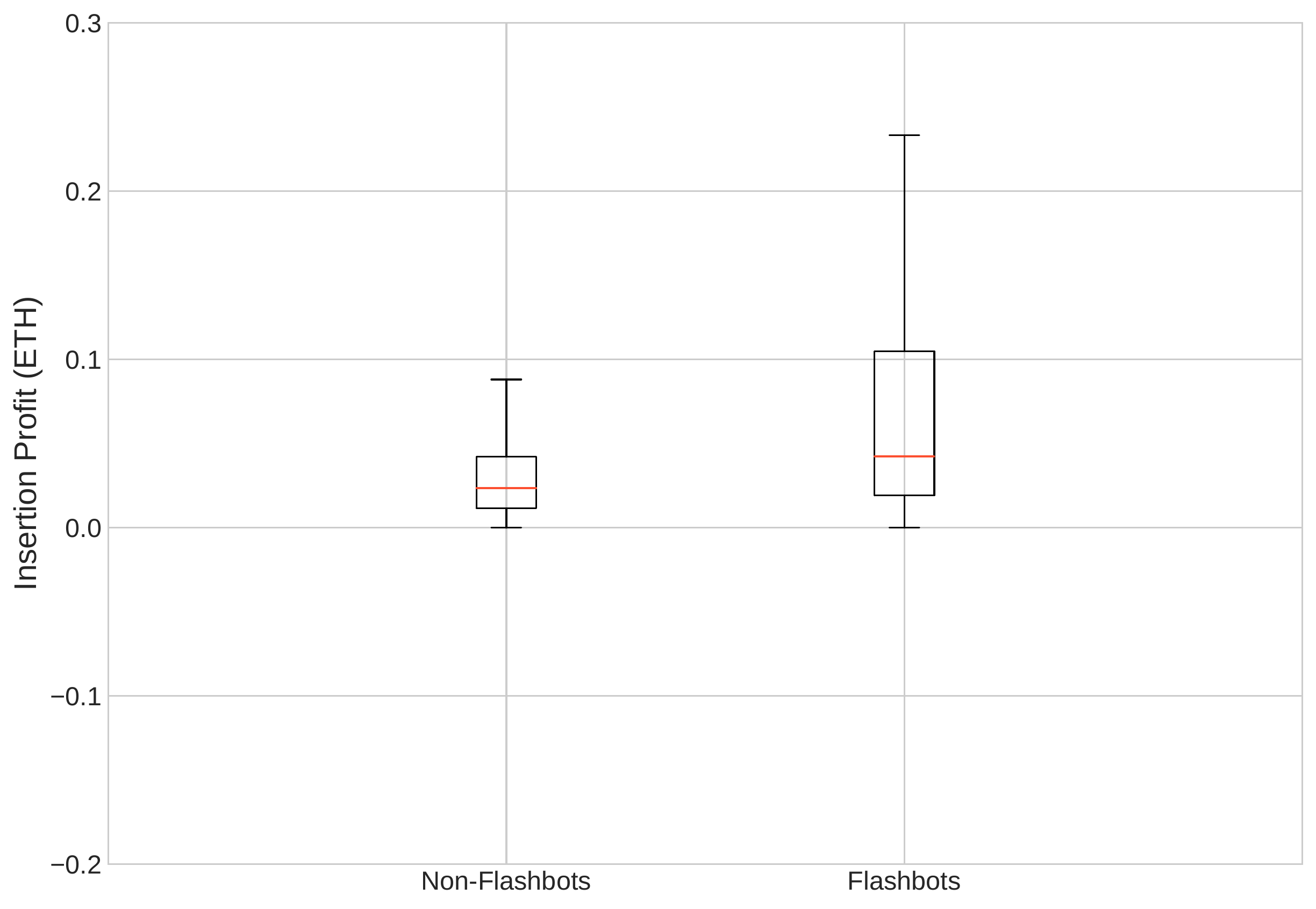}%
      \caption{Miners}%
      \label{fig:miner_insertion_profits}%
    \end{subfigure}%

    \begin{subfigure}[t]{\columnwidth}%
      \centering
      \includegraphics[width=0.8\columnwidth]{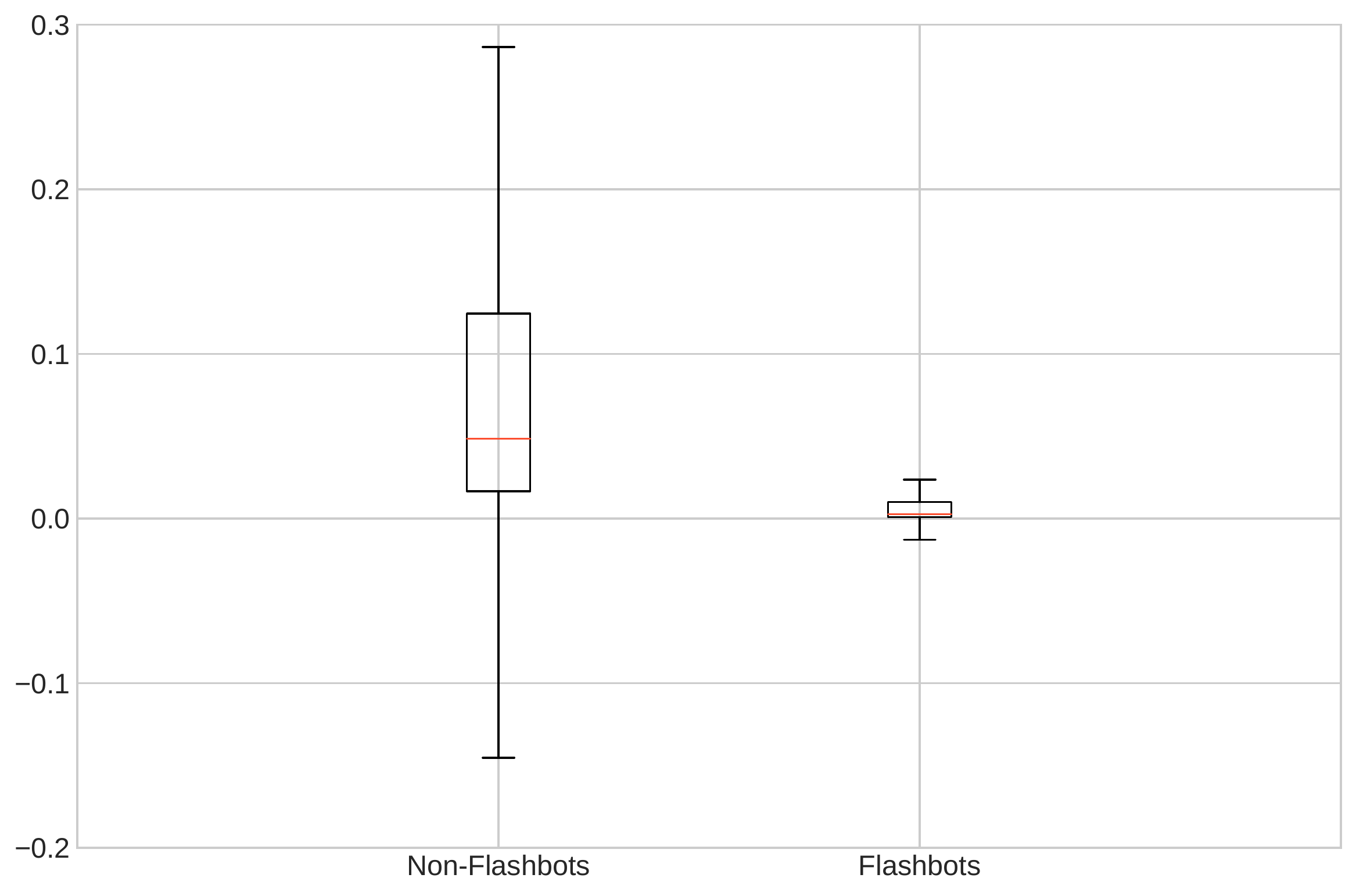}%
      \caption{Non-Miners (Searchers)}%
      \label{fig:non_miner_insertion_profits}%
    \end{subfigure}%
    \caption{Sandwich profits (ETH) for different subpopulations. Horizontal
      bars represent the median profit.}%
    \label{fig:insertion_profits}%
\end{figure}

We show, in \Cref{fig:miner_insertion_profits}, that miners are in fact seeing
slightly higher average profits on sandwich frontrunning when using
\fb than without \fb (\SI{0.125}{ETH} and \SI{0.048}{ETH}, respectively). This comes at the cost of somewhat increased
variance (standard deviation of \num{0.415} and \num{0.127}, respectively). As the average
profit is the \emph{expected} profit, a rational miner will prefer to use \fb.

Conversely, non-miners (or in the \fb case: \emph{searchers}) see
notably lower average profit when using \fb: \SI{0.02}{ETH}, compared to
\SI{0.13}{ETH} without \fb. Profit from non-\fb usage does, however, come with
lower standard deviations, \num{0.154} versus \num{0.560}. This means \fb is a more reliable income source---a important
attribute for some users. However, this reliability is offset by lower expected
income. Since miners are making \SI{260}{\percent} more profit and
non-miners are making \SI{84.4}{\percent} less, it does not make rational sense
for most non-miners to continue their sandwich MEVs in \fb.

In \Cref{fig:non_miner_insertion_profits}, we see that non-miners have significant
probability of a sandwich MEV incurring a loss. We discuss this further in
\Cref{sec:neg_prof}.

\subsection{Negative Profits}\label{sec:neg_prof}
A number of transactions in the \fb epoch yielded negative profit for
searchers which should not happen. A searcher standing to make negative profit
would not forward such a bundle to a miner. Here we explore those transactions
in depth.

In the \fb epoch, there have been \num{7666} unprofitable MEVs out of a total of
\num{485680} transactions. About \SI{1.58}{\percent}. Unprofitable transactions in \fb total
\SI{113.67}{ETH} or roughly \SI{378399.40}{USD}.

The cause of these negative profits is faulty contracts provided by the seachers
in their submitted bundles.
Integrating logical checks of correctness is difficult in any distributed
system, no less for one as volatile as Ethereum.
This is problematic in the context of the \fb ethos, as \fb is billed as a
system to assist those without the resources to engineer their own involvement
in MEV---it does not protect these low resource searchers from taking
on losses due to unprofitable transactions. These occurrences do not well support
\Cref{goal:democ_mev_extract} (democratizing MEV extraction)\cameraAdd{, and
  constitute a real risk for searchers}.

\section{Private MEV Extraction}\label{sec:private}

\fb is not the only private pool available to non-miners and
miners. 
Other private pools have
been proposed such as the Taichi Network~\cite{taichi} and the Eden
Network~\cite{eden}---the Eden Network is currently
still active, but the Taichi Network is defunct as of October 15th,
2021~\cite{taichi}.
Along with well known private pools, miners can collaborate to make their own
private pools, and even participate in multiple private pools concurrently.
This raises the question of whether \fb is the most prominent private pool and
whether miners only use \fb or if they also use other private pools
concurrently.

Similar to \fb, both Taichi and Eden enable users to submit their private
transactions to the network via a dedicated RPC endpoint, where a trusted miner
will include these private transactions within a future block. Unlike \fb,
however, neither network publicly discloses which private transactions were
relayed via their network and mined by their miners. Currently known techniques
are insufficient to identify how much MEV has been privately extracted using
Taichi or Eden. We therefore only differentiate between public MEV extractions,
\fb MEV extractions, and non-\fb private extractions MEV.

In this section, we investigate how much of the privately extracted MEV is due
to \fb and how \fb contributes to the number of private transactions currently
being mined in Ethereum.

\vspace{-5pt}
\subsection{Discovering Private MEV Extractions}\label{sec:priv_mev_disc}
The Ethereum blockchain does not include any explicit indication of whether a
transaction was public or private. We next describe our method for inferring this
information. At a high-level, we calculated the intersection between the set of
publicly observed pending transactions and the set of transactions included in
blocks on the blockchain---the transactions \emph{not} in the intersection are,
by definition, private. This assumes that our node saw the vast majority of
transactions propagated through the network---an assumption consistent with
previous measurements~\cite{kiffer2021gosssip,pietExtractingGodlSic2022}.

We did this intersection calculation for blocks ranging from block number
\num{13670000} (\ie November 23rd, 2021) to \num{14444725} (\ie March 23rd,
2022). This range aligns with the window of our pending transactions data
collection, and resulted in an analysis of \num{774725} blocks---about 4 months
of data.

Next, we identified the subset of private transactions which were MEV
extractions. We did so by a similar technique as in \Cref{sec:sand_detect}.
The difference is that we only searched for \emph{MEVs} in the private transaction
subset, while we only searched for \emph{victims} in the public transaction set. This is
because frontrunning other \fb transactions is disallowed in \fb and
frontrunning other private pool transactions is not possible.

For example, sandwiches are always composed of three transactions: two
transactions created by the MEV extractor, and one transaction created by the
victim. The victim's transaction is ordered between the two transactions
of the MEV extractor.
Thus, to identify private sandwiches, we simply check whether both the first
transaction and the third transaction are \emph{not} part of our dataset of publicly
observed pending transactions, while the second transaction \emph{is} included
in our dataset of publicly observed pending transactions.

\subsection{Private MEV Distribution}\label{sec:priv_dist}

\begin{figure}%
    \centering%
    \includegraphics[width=\columnwidth]{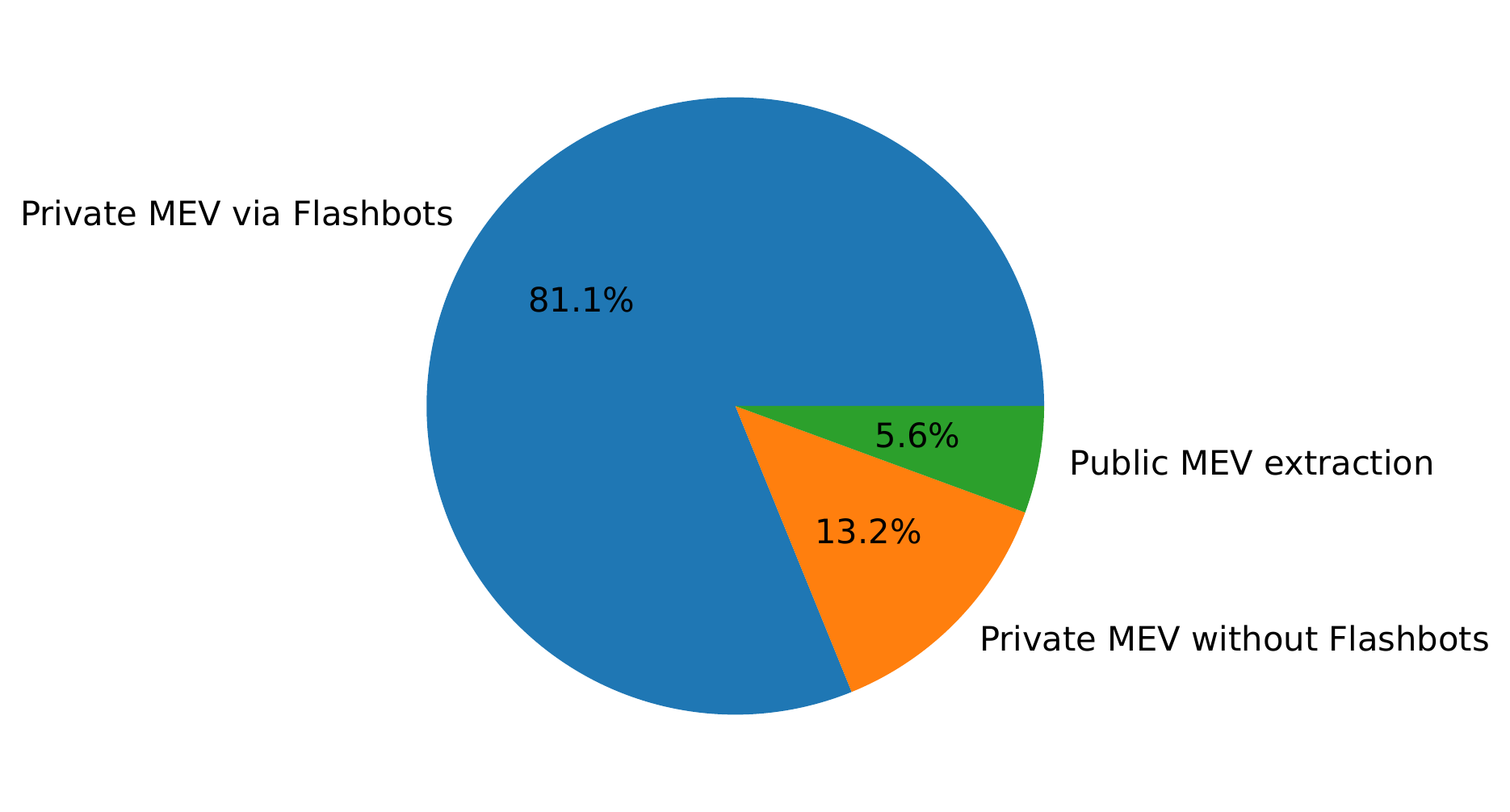}
    \caption{Distribution of private vs. public MEV extraction.}%
    \label{fig:private_mev}%
\end{figure}

Like \fb transactions and publicly propagated transactions, private transactions
can be of any known type of MEV extraction. Analyzing the private transactions
with this in mind, we found \num{80093} (\SI{10.34}{\percent}) blocks containing at least one sandwich MEV within
the \num{774725} blocks we analyzed, In total, we found \num{99928} sandwiches
between November 23rd, 2021 and March 23rd, 2022. Out of these \num{99928}
sandwiches, \num{81089} (\SI{81.15}{\percent}) were performed using
\fb and the remaining \num{18839} (\SI{18.85}{\percent}) were performed outside of \fb.
Thus, the majority of the sandwich attacks between November 23rd, 2021 and March
23rd, 2022, were performed by using \fb.

When applying the heuristics described in \Cref{sec:priv_mev_disc} to detect
private MEV extraction on sandwiches, we find that out of the \num{18839}
sandwiches that were not performed using \fb, \num{13238} (\SI{70.27}{\percent})
were private. This means that only \num{5601} sandwich attacks were public. In
other words, from the overall \num{99928} sandwich attacks only
\SI{5.6}{\percent} were carried out using the public mempool. This shows that
private pools such as \fb are very popular for MEV extraction and that \fb is
currently the dominating private pool. However, our results also show that other
private pools seem to coexist and that miners engage in private
MEV extraction.

\subsection{Private Miner MEV Extraction}

Miners and non-miners alike have flexibility in deciding which private pools, if
any, they want to join. As we've argued, private pools can be lucrative in some
cases, so here we evaluate the prevalence of miners and non-miners seeking such
opportunities though non-\fb private pools. We identified miner addresses and
account addresses that performed private non-\fb MEV sandwiches to
better understand whether miners are engaging in MEV extraction themselves or
whether they are part of broader private pools.

We found a total of 35 different miner addresses that mined private non-\fb
MEV sandwiches and 41 account addresses that performed private non-\fb
MEV sandwiches. We started by counting, for each account address, the
number of unique miner addresses that mined an accounts address's private
non-\fb sandwich MEV. The intuition here is that if a miner engaged in
private MEV extraction then we should find an account address whose private
non-\fb sandwich MEVs have only ever been mined by a single
miner. It is very unlikely that all private MEV transactions would have been
mined by the same miner within a private pool with multiple miners. Thus,
if we find that an account address's private non-\fb sandwich MEV
were mined by multiple miner addresses, then we can assume
that these were mined via a private pool where these miners are participants.

We found two account addresses where each address's private transactions were only
mined by a single miner, thus two miners that are likely engaging in private
MEV extraction on their own. The first account address is
\ttt{0x42B2C65dB7F9e3b6c26Bc6151CCf30CcE0fb99EA} with 30 private non-\fb
sandwiches, all mined by the miner address
\ttt{0x7F101fE45e6649A6fB8F3F8B43ed03D353f2B90c}, which according to
Etherscan~\cite{etherscan}, is a part of
Flexpool~\cite{flexpool}. The second account address is
\ttt{0xDD28D64E40e00aF54a0B514753 9A515C4A0bC1c5} with 121 private non-\fb
sandwiches, all mined by the miner address
\ttt{0x829BD824B016326A401d083B33D092293333A830}, which again according to Etherscan,
is part of F2Pool~\cite{f2pool}. Since we did not see
other account addresses engaging with only these two miners, we can eliminate the
possibility that each of those two miners provide their own private pool for
users. We also note that both miners have mined private transactions of other
account addresses that have engaged with other miners. Thus, both Flexpool and
F2Pool not only perform private MEV extraction on their own, but they also
participate in other private pools besides \fb.

\section{Related Work}
\label{sec:related_work}

\citet{EskandariMC19} were the first to propose a taxonomy of frontrunning
attacks on the blockchain. However, \citet{daian2020flash} were the
first to introduce the term ``MEV'' and to research frontrunning attacks from an
economical perspective. They did so by monitoring the mempool and studying
\emph{priority gas auctions} (PGAs) in realtime.
As a response to the negative effects of PGAs and the sharp increase of frontrunning
attacks, projects such as \fb~\cite{flashbotsdocs} and the Eden
Network~\cite{eden} have been proposed. These projects form private transaction
pools which reflect private agreements with miners that allows users to bypass
the public mempool and submit transactions privately.

While \citet{EskandariMC19} provide a taxonomy and \citet{daian2020flash} show
the existence of frontrunning in the wild, \citet{ferreira2021frontrunner} were
the first to quantify past frontrunning attacks using historical blockchain
data. \citet{qinQuantifyingBlockchainExtractable2021} on the other hand focused
on quantifying MEV extraction.
Both works discover large increases in
frontrunning attacks over recent years. Our work leverages and extends their
techniques in order to measure how much MEV is being extracted using private
pools such as \fb. \citet{qinQuantifyingBlockchainExtractable2021} also provide a
theoretical analysis of network congestion in the presence of private pools, and conclude that \fb does not reduce the
such congestion. While this may be true, our work shows that \fb does at least
reduce gas prices in some instances.

Recent works by \citet{pietExtractingGodlSic2022} and
\citet{capponi2022evolution} analyze the profit distribution within \fb and
conclude, similarly to our work, that miners are earning most of the profit.
However, neither work analyzes if this is unique to \fb or if this was
already the case before \fb. Our work measures MEV extraction before and after
the inception of \fb and concludes that searchers were making more profit prior
to \fb. Our results also show that the number of searchers using \fb is
decreasing. Moreover, while \citet{pietExtractingGodlSic2022} focuses on
measuring the distribution of private transactions, our work focuses on
measuring the existence and distribution of private pools.

Numerous other frontrunning countermeasures have also been proposed.
One strategy is to prevent frontrunning at the application layer either through
an advanced commit-and-reveal scheme~\cite{breidenbach2018enter}, tighter
slippage protection~\cite{heimbach2022eliminating}, or a more
frontrunning-resistant design of
DEXes~\cite{bentov2019tesseract,zhou2021a2mm,ciampi2021fairmm,baum2021p2dex,mcmenamin2022fairtradex}.
The main issue with these solutions is that they introduce new costs and
each protects against only a single type of frontrunning attack.
Another strategy is to prevent frontrunning at the
consensus layer by proposing a fair ordering of
transactions~\cite{kelkar2020order,kelkar2021themis,kursawe2020wendy} or
introducing transaction privacy~\cite{kokoris2018calypso,yakira2021helix}.
Unfortunately, none of these techniques are broadly adopted as they are not
applicable to large public blockchains such as Ethereum.


\section{Concluding Discussion}\label{sec:conclusion}

The rise of \fb indicates that many Ethereum users see value in what
\fb offers. The public charter of \fb is laudable, but prior to this work, it
was not clear if its goals were being met, and thus if it was a viable,
long-term solution.
This work yields the following three takeaways.
\begin{enumerate}
  \item \fb does not provide a one-stop, easy-access tool for MEV extraction. Users
        are still required to have the know-how to design MEV extraction
        solutions.
  \item \fb has increased the total number of MEV transactions in Ethereum, while
        reducing gas prices. Lower gas prices correspond to lower transaction
        fees for users.
  \item \fb has not been equally useful to all users. Miners have profited much more than searchers.
\end{enumerate}

\subsection{Mission Accomplished?}
Despite \fb' efforts towards its goals, we must question how much progress it has
actually made in achieving those goals. And if has made enough progress to be
considered a success.

The first goal (\Cref{goal:illuminate}: Illuminating MEV) is to increase transparency of MEV in
the mempool. If \fb has spurred more users to engage with other
private pools then transparency cannot be said to be sufficiently addressed. It
is not clear that, indeed, \fb was the driving factor in users moving towards other
private pools, but private pools have certainly replaced public MEV to a large
extent (\Cref{sec:priv_dist}).

Despite our efforts to contact the \fb project with the hope of measuring,
directly, their pool of pending transactions, we were unable to get through to
them. This is a strike against \fb as in this regard, it is no different than
any other ``dark'', private pool. Conversely, their measurement
dashboard~\cite{transparency} has many interesting plots and datapoints, though
it contains no direct analysis of \fb' real world use cases.
Another concern is that the methodology for generating these plots is not
sufficiently open, making comparison difficult---if not impossible. Overall, there
is no clear answer to whether \Cref{goal:illuminate} has been achieved. \fb is
much more open than other private pools, but does not seem to be an optimal
solution in illuminating MEV behavior.

The second goal of \fb (\Cref{goal:democ_mev_extract}: Democratizing MEV) is
similarly complex. Through \fb, more people have
access to MEV-permitting infrastructure than ever before. However, this comes at
the risk of taking on losses (\Cref{sec:neg_prof}), especially for
non-miners---who already have fewer resources than their miner counterparts.
This is because some degree of knowledge in writing smart contracts and understanding MEV opportunities is required to use \fb. This is beneficial for whoever can afford the effort to
write (and verify) such contracts, but leaves non-experts in the same position
as before. In fact, they are in a worse position, because now their are many
more users who will be trying to frontrun them. Again, there is no clear answer
to whether MEV is democratized, so we stick to the same refrain: more than
before, but less than it could be.

The third goal of \fb is to distribute the benefits of MEV extraction
(\Cref{goal:dist_bene}: Distributing Benefits). On this point, we note that after an initially steep
climb in usage, \fb has seen a decline in the number of blocks published
(see \Cref{fig:fb_hashpower}). If rational actors are declining in their usage of \fb,
then it can only be assumed that they are not finding it as profitable as other
options. This is unsurprising given that non-miners (a sizeable population) are
seeing much less profit than they were before \fb
(see \Cref{fig:non_miner_insertion_profits}). And so for those users no longer using
\fb, the MEV problem has not been addressed. On this count, it is clear that \fb
is not achieving its goal. Benefits are even more skewed than before.

\subsection{Effects on Ethereum}
MEV extraction does not happen in a vacuum. Their are externalities affecting
the entire Ethereum ecosystem. On one hand, \fb has driven down gas prices
(see \Cref{fig:gas_price_txs_per_day}), which seems to be a good thing. We
conjecture that the reduction in gas prices is because \fb logically splits the
mempool into two disjoint pools, one for the MEV extractors and one for
everyone else. Thus two different gas price auctions are occurring, the
competition on one pool does not impact the other pool. This is contrary to
what was previously happening where non-MEV extracting users had to pay more
because of \emph{priorty gas auctions}~\cite{daian2020flash} happening in the
public mempool.

On the other hand, fees for MEV extractors, have risen drastically. \fb has made
it difficult for searchers to optimize bids, because of its sealed bid
auctions. 
Searchers do not see the bids of other searchers, and so to increase their
chances of extracting MEV, they prefer to increase their miner tip in order to be
selected by the miner. This design choice solely benefits miners and not
searchers.
Perhaps another type of auction would be better. This gas-price tradeoff has come out
squarely in favor of miners who now indirectly force searchers to pay higher fees---a
tragedy of the commons.

\subsection{Are We Going in the Right Direction?}
We now question whether or not the goals of \fb are good goals, per se. In
traditional financial markets, frontrunning is considered predatory
behavior~\cite{Manahov_2016}. Many agree that it is equally pernicious in the
context of blockchain economies~\cite{Felten_2020, dark_forest_2020}. However, unlike
other works that attempt to disallow frontrunning through cryptographic
means \cite{kelkar2021themis, kelkar2020order, kursawe2020wendy, chainlinkblog},
\fb actually makes MEV extraction easier. Their reasoning being that MEV
extraction is going to happen anyway, so it might as well be extracted in a way
that is more broadly beneficial.

This view has been called into question. \citet{juels2021Apr} directly call \fb,
and other frontrunning-as-a-service (FaaS) solutions, ``theft''.
\citet{Felten_2020} describes \fb-like solutions as ``a poorly designed tax on
users.'' Arguing that the additional income for searchers is coming at the cost
of others in the Ethereum ecosystem (\ie miners).

It is no surprise that so many miners have joined \fb. The ``London''
hard fork resulted in a large reduction of miners' revenue through block
rewards~\cite{eip1559analysis}. \fb' auction system seems to be designed with
miner revenue in mind and thus makes it perfect for miners to compensate for the
lost revenue due to the hard fork.

As discussed in \Cref{sec:related_work}, a number of technical solutions have been proposed to mitigate or reduce MEV extraction via frontrunning. \citet{pietExtractingGodlSic2022} suggest changing the
Ethereum protocol to randomly order transactions---while this is certainly
possible using a random seed derived from the previous block, it has other
critical flaws. Consider a sandwich MEV, which requires three
transactions to appear in order. On average, after a random shuffle, the victim's
transaction will appear in the middle of the block. Then, with \SI{50}{\percent}
probability, the first transaction in the sandwich will appear \emph{before} the victim,
and with \SI{50}{\percent} probability, the third transaction in the sandwich will appear
\emph{after} the victim. This means, that despite the shuffle, there is still a
\SI{25}{\percent} chance that the MEV will take place. This probability doubles
if we consider arbitrage or liquidation MEVs which only require a single
frontrun or backrun. The probability of an MEV extractor could be further
increased by simply including more transactions---essentially throwing darts,
and hoping one sticks. \cameraAdd{This technique would certainly lower the
  likelihood of MEV success, but for cases with large enough payouts, the
  expected income would still be positive and thus the behavior is likely
  to continue.} For this reason, we do not consider randomization viable.

Better approaches such as in \cite{kelkar2021themis,kelkar2020order} offer
cryptographic guarantees along with a definition of \emph{fair-ordering} bespoke
to the context of blockchains. The limitation of these approaches being that
they are currently only applicable to private blockchains where the number of
nodes is small and reliable as opposed to the reality of public blockchains like
Ethereum.

\section{Ethics}

\cameraAdd{These measurements concern an actively used financial system. This
  system is operated by people who may take our analyses into account in their
  usage of (or abstention from) \fb and, more broadly, Ethereum.}
However, our work exclusively collected and parsed public data that was voluntarily
distributed by the originators of that data. All data collected is in this work is
considered publicly available by the protocol they operate within. Additionally,
none of the analysis or collection focused on individuals, analysis was only
done via aggregations.

\section{Acknowledgments}

We would like to thank our anonymous reviewers and our shepherd
Michael Sirivianos for their valuable comments and feedback. 
This work was partly supported by the Ripple University
Blockchain Research Initiative (UBRI) grant number 2018-188548 (5855).

\bibliographystyle{meta/ACM-Reference-Format}
\bibliography{main}


\begin{thebibliography}{58}


\ifx \showCODEN    \undefined \def \showCODEN     #1{\unskip}     \fi
\ifx \showDOI      \undefined \def \showDOI       #1{#1}\fi
\ifx \showISBNx    \undefined \def \showISBNx     #1{\unskip}     \fi
\ifx \showISBNxiii \undefined \def \showISBNxiii  #1{\unskip}     \fi
\ifx \showISSN     \undefined \def \showISSN      #1{\unskip}     \fi
\ifx \showLCCN     \undefined \def \showLCCN      #1{\unskip}     \fi
\ifx \shownote     \undefined \def \shownote      #1{#1}          \fi
\ifx \showarticletitle \undefined \def \showarticletitle #1{#1}   \fi
\ifx \showURL      \undefined \def \showURL       {\relax}        \fi
\providecommand\bibfield[2]{#2}
\providecommand\bibinfo[2]{#2}
\providecommand\natexlab[1]{#1}
\providecommand\showeprint[2][]{arXiv:#2}

\bibitem[0x(2016)]%
        {0x}
 \bibinfo{year}{2016}\natexlab{}.
\newblock
\newblock
\urldef\tempurl%
\url{https://www.0x.org}
\showURL{%
\tempurl}


\bibitem[ban(2017)]%
        {bancor}
 \bibinfo{year}{2017}\natexlab{}.
\newblock
\newblock
\urldef\tempurl%
\url{https://home.bancor.network}
\showURL{%
\tempurl}


\bibitem[uni(2018)]%
        {uniswap}
 \bibinfo{year}{2018}\natexlab{}.
\newblock
\newblock
\urldef\tempurl%
\url{https://uniswap.org}
\showURL{%
\tempurl}


\bibitem[com(2019)]%
        {compound}
 \bibinfo{year}{2019}\natexlab{}.
\newblock
\newblock
\urldef\tempurl%
\url{https://compound.finance}
\showURL{%
\tempurl}


\bibitem[bal(2019)]%
        {balancer}
 \bibinfo{year}{2019}\natexlab{}.
\newblock
\newblock
\urldef\tempurl%
\url{https://balancer.fi}
\showURL{%
\tempurl}


\bibitem[dyd(2019)]%
        {dydx}
 \bibinfo{year}{2019}\natexlab{}.
\newblock
\newblock
\urldef\tempurl%
\url{https://dydx.exchange}
\showURL{%
\tempurl}


\bibitem[aav(2020)]%
        {aave}
 \bibinfo{year}{2020}\natexlab{}.
\newblock
\newblock
\urldef\tempurl%
\url{https://aave.com/}
\showURL{%
\tempurl}


\bibitem[sus(2020)]%
        {sushiswap}
 \bibinfo{year}{2020}\natexlab{}.
\newblock
\newblock
\urldef\tempurl%
\url{https://www.sushi.com}
\showURL{%
\tempurl}


\bibitem[cur(2020)]%
        {curve}
 \bibinfo{year}{2020}\natexlab{}.
\newblock
\newblock
\urldef\tempurl%
\url{https://curve.fi}
\showURL{%
\tempurl}


\bibitem[tai(2020)]%
        {taichi}
 \bibinfo{year}{2020}\natexlab{}.
\newblock
\newblock
\urldef\tempurl%
\url{https://github.com/Taichi-Network/docs}
\showURL{%
\tempurl}


\bibitem[dar(2020)]%
        {dark_forest_2020}
 \bibinfo{year}{2020}\natexlab{}.
\newblock
\newblock
\urldef\tempurl%
\url{https://www.paradigm.xyz/2020/08/ethereum-is-a-dark-forest}
\showURL{%
\tempurl}


\bibitem[cha(2021)]%
        {chainlinkblog}
 \bibinfo{year}{2021}\natexlab{}.
\newblock \bibinfo{title}{{Fair Sequencing Services: Enabling a Provably Fair
  DeFi Ecosystem}}.
\newblock
\newblock
\urldef\tempurl%
\url{https://blog.chain.link/chainlink-fair-sequencing-services-enabling-a-provably-fair-defi-ecosystem}
\showURL{%
\tempurl}
\newblock
\shownote{[Online; accessed 23. Nov. 2021]}.


\bibitem[fla(2021)]%
        {flashbotsdocs}
 \bibinfo{year}{2021}\natexlab{}.
\newblock \bibinfo{title}{{Overview {$\vert$} Flashbots Docs}}.
\newblock
\newblock
\urldef\tempurl%
\url{https://docs.flashbots.net/flashbots-auction/overview}
\showURL{%
\tempurl}
\newblock
\shownote{[Online; accessed 23. Nov. 2021]}.


\bibitem[mev(2022a)]%
        {mev-geth}
 \bibinfo{year}{2022}\natexlab{a}.
\newblock
\newblock
\urldef\tempurl%
\url{https://github.com/flashbots/mev-geth}
\showURL{%
\tempurl}


\bibitem[mev(2022b)]%
        {mev-inspect}
 \bibinfo{year}{2022}\natexlab{b}.
\newblock
\newblock
\urldef\tempurl%
\url{https://github.com/flashbots/mev-inspect-rs}
\showURL{%
\tempurl}


\bibitem[go-(2022)]%
        {go-ethereum}
 \bibinfo{year}{2022}\natexlab{}.
\newblock
\newblock
\urldef\tempurl%
\url{https://github.com/ethereum/go-ethereum}
\showURL{%
\tempurl}


\bibitem[tra(2022)]%
        {transparency}
 \bibinfo{year}{2022}\natexlab{}.
\newblock
\newblock
\urldef\tempurl%
\url{https://dashboard.flashbots.net/}
\showURL{%
\tempurl}


\bibitem[mev(2022c)]%
        {mev-explore}
 \bibinfo{year}{2022}\natexlab{c}.
\newblock
\newblock
\urldef\tempurl%
\url{https://explore.flashbots.net/}
\showURL{%
\tempurl}


\bibitem[coi(2022a)]%
        {coingecko}
 \bibinfo{year}{2022}\natexlab{a}.
\newblock \bibinfo{title}{{CoinGecko}}.
\newblock
\newblock
\urldef\tempurl%
\url{https://www.coingecko.com/en/api}
\showURL{%
\tempurl}
\newblock
\shownote{[Online; accessed 18. May. 2022]}.


\bibitem[coi(2022b)]%
        {coinmarketcap}
 \bibinfo{year}{2022}\natexlab{b}.
\newblock \bibinfo{title}{{CoinMarketCap}}.
\newblock
\newblock
\urldef\tempurl%
\url{https://coinmarketcap.com/}
\showURL{%
\tempurl}
\newblock
\shownote{[Online; accessed 18. May. 2022]}.


\bibitem[eth(2022)]%
        {etherscan}
 \bibinfo{year}{2022}\natexlab{}.
\newblock \bibinfo{title}{{Etherscan}}.
\newblock
\newblock
\urldef\tempurl%
\url{https://etherscan.io}
\showURL{%
\tempurl}
\newblock
\shownote{[Online; accessed 18. May. 2022]}.


\bibitem[f2p(2022)]%
        {f2pool}
 \bibinfo{year}{2022}\natexlab{}.
\newblock \bibinfo{title}{f2pool}.
\newblock
\newblock
\urldef\tempurl%
\url{https://www.f2pool.com}
\showURL{%
\tempurl}
\newblock
\shownote{[Online; accessed 18. May. 2022]}.


\bibitem[fle(2022)]%
        {flexpool}
 \bibinfo{year}{2022}\natexlab{}.
\newblock \bibinfo{title}{{Flexpool.io}}.
\newblock
\newblock
\urldef\tempurl%
\url{https://www.flexpool.io/}
\showURL{%
\tempurl}
\newblock
\shownote{[Online; accessed 18. May. 2022]}.


\bibitem[ukr(2022)]%
        {ukraineCrypto}
 \bibinfo{year}{2022}\natexlab{}.
\newblock \bibinfo{title}{Live Updates: Ukraine Government Turns to Crypto to
  Crowdfund Millions of Dollars}.
\newblock
\newblock
\urldef\tempurl%
\url{https://www.elliptic.co/blog/live-updates-millions-in-crypto-crowdfunded-for-the-ukrainian-military}
\showURL{%
\tempurl}
\newblock
\shownote{[Online; accessed 02. September. 2022]}.


\bibitem[fb_(2022)]%
        {fb_block_api}
 \bibinfo{year}{2022}\natexlab{}.
\newblock \bibinfo{title}{{mev-blocks}}.
\newblock
\newblock
\urldef\tempurl%
\url{https://blocks.flashbots.net}
\showURL{%
\tempurl}
\newblock
\shownote{[Online; accessed 18. May. 2022]}.


\bibitem[web(2022)]%
        {web3js}
 \bibinfo{year}{2022}\natexlab{}.
\newblock \bibinfo{title}{{web3.js - Ethereum JavaScript API}}.
\newblock
\newblock
\urldef\tempurl%
\url{https://web3js.readthedocs.io}
\showURL{%
\tempurl}
\newblock
\shownote{[Online; accessed 18. May. 2022]}.


\bibitem[Baum et~al\mbox{.}(2021)]%
        {baum2021p2dex}
\bibfield{author}{\bibinfo{person}{Carsten Baum}, \bibinfo{person}{Bernardo
  David}, {and} \bibinfo{person}{Tore~Kasper Frederiksen}.}
  \bibinfo{year}{2021}\natexlab{}.
\newblock \showarticletitle{P2DEX: privacy-preserving decentralized
  cryptocurrency exchange}. In \bibinfo{booktitle}{\emph{International
  Conference on Applied Cryptography and Network Security}}. Springer,
  \bibinfo{pages}{163--194}.
\newblock


\bibitem[Bentov et~al\mbox{.}(2019)]%
        {bentov2019tesseract}
\bibfield{author}{\bibinfo{person}{Iddo Bentov}, \bibinfo{person}{Yan Ji},
  \bibinfo{person}{Fan Zhang}, \bibinfo{person}{Lorenz Breidenbach},
  \bibinfo{person}{Philip Daian}, {and} \bibinfo{person}{Ari Juels}.}
  \bibinfo{year}{2019}\natexlab{}.
\newblock \showarticletitle{Tesseract: Real-time cryptocurrency exchange using
  trusted hardware}. In \bibinfo{booktitle}{\emph{Proceedings of the 2019 ACM
  SIGSAC Conference on Computer and Communications Security}}.
  \bibinfo{pages}{1521--1538}.
\newblock


\bibitem[Bernhardt and Taub(2008)]%
        {Bernhardt_Taub_2008}
\bibfield{author}{\bibinfo{person}{Dan Bernhardt} {and} \bibinfo{person}{Bart
  Taub}.} \bibinfo{year}{2008}\natexlab{}.
\newblock \showarticletitle{Front-running dynamics}.
\newblock \bibinfo{journal}{\emph{Journal of Economic Theory}}
  \bibinfo{volume}{138}, \bibinfo{number}{1} (\bibinfo{date}{Jan}
  \bibinfo{year}{2008}), \bibinfo{pages}{288–296}.
\newblock
\showISSN{00220531}
\urldef\tempurl%
\url{https://doi.org/10.1016/j.jet.2007.05.005}
\showDOI{\tempurl}


\bibitem[Breidenbach et~al\mbox{.}(2018)]%
        {breidenbach2018enter}
\bibfield{author}{\bibinfo{person}{Lorenz Breidenbach}, \bibinfo{person}{Phil
  Daian}, \bibinfo{person}{Florian Tram{\`e}r}, {and} \bibinfo{person}{Ari
  Juels}.} \bibinfo{year}{2018}\natexlab{}.
\newblock \showarticletitle{Enter the Hydra: Towards Principled Bug Bounties
  and $\{$Exploit-Resistant$\}$ Smart Contracts}. In
  \bibinfo{booktitle}{\emph{27th USENIX Security Symposium (USENIX Security
  18)}}. \bibinfo{pages}{1335--1352}.
\newblock


\bibitem[Capponi et~al\mbox{.}(2022)]%
        {capponi2022evolution}
\bibfield{author}{\bibinfo{person}{Agostino Capponi}, \bibinfo{person}{Ruizhe
  Jia}, {and} \bibinfo{person}{Ye Wang}.} \bibinfo{year}{2022}\natexlab{}.
\newblock \showarticletitle{The Evolution of Blockchain: from Lit to Dark}.
\newblock \bibinfo{journal}{\emph{arXiv preprint arXiv:2202.05779}}
  (\bibinfo{year}{2022}).
\newblock


\bibitem[Ciampi et~al\mbox{.}(2021)]%
        {ciampi2021fairmm}
\bibfield{author}{\bibinfo{person}{Michele Ciampi}, \bibinfo{person}{Muhammad
  Ishaq}, \bibinfo{person}{Malik Magdon-Ismail}, \bibinfo{person}{Rafail
  Ostrovsky}, {and} \bibinfo{person}{Vassilis Zikas}.}
  \bibinfo{year}{2021}\natexlab{}.
\newblock \showarticletitle{FairMM: A fast and frontrunning-resistant crypto
  market-maker}.
\newblock \bibinfo{journal}{\emph{Cryptology ePrint Archive}}
  (\bibinfo{year}{2021}).
\newblock


\bibitem[Daian et~al\mbox{.}(2020)]%
        {daian2020flash}
\bibfield{author}{\bibinfo{person}{Philip Daian}, \bibinfo{person}{Steven
  Goldfeder}, \bibinfo{person}{Tyler Kell}, \bibinfo{person}{Yunqi Li},
  \bibinfo{person}{Xueyuan Zhao}, \bibinfo{person}{Iddo Bentov},
  \bibinfo{person}{Lorenz Breidenbach}, {and} \bibinfo{person}{Ari Juels}.}
  \bibinfo{year}{2020}\natexlab{}.
\newblock \showarticletitle{Flash boys 2.0: Frontrunning in decentralized
  exchanges, miner extractable value, and consensus instability}. In
  \bibinfo{booktitle}{\emph{2020 IEEE Symposium on Security and Privacy (SP)}}.
  IEEE, \bibinfo{pages}{910--927}.
\newblock


\bibitem[Eskandari et~al\mbox{.}(2019)]%
        {EskandariMC19}
\bibfield{author}{\bibinfo{person}{Shayan Eskandari},
  \bibinfo{person}{Seyedehmahsa Moosavi}, {and} \bibinfo{person}{Jeremy
  Clark}.} \bibinfo{year}{2019}\natexlab{}.
\newblock \showarticletitle{SoK: Transparent Dishonesty: Front-Running Attacks
  on Blockchain}. In \bibinfo{booktitle}{\emph{Financial Cryptography and Data
  Security - {FC} 2019 International Workshops, {VOTING} and WTSC, St. Kitts,
  St. Kitts and Nevis, February 18-22, 2019, Revised Selected Papers}}
  \emph{(\bibinfo{series}{Lecture Notes in Computer Science},
  Vol.~\bibinfo{volume}{11599})}, \bibfield{editor}{\bibinfo{person}{Andrea
  Bracciali}, \bibinfo{person}{Jeremy Clark}, \bibinfo{person}{Federico
  Pintore}, \bibinfo{person}{Peter~B. R{\o}nne}, {and}
  \bibinfo{person}{Massimiliano Sala}} (Eds.). \bibinfo{publisher}{Springer},
  \bibinfo{pages}{170--189}.
\newblock


\bibitem[Felten(2020)]%
        {Felten_2020}
\bibfield{author}{\bibinfo{person}{Ed Felten}.}
  \bibinfo{year}{2020}\natexlab{}.
\newblock \bibinfo{title}{MEV auctions considered harmful}.
\newblock
\newblock
\urldef\tempurl%
\url{https://medium.com/offchainlabs/mev-auctions-considered-harmful-fa72f61a40ea}
\showURL{%
\tempurl}


\bibitem[Gencer et~al\mbox{.}(2018)]%
        {gencer2018decentralization}
\bibfield{author}{\bibinfo{person}{Adem~Efe Gencer}, \bibinfo{person}{Soumya
  Basu}, \bibinfo{person}{Ittay Eyal}, \bibinfo{person}{Robbert van Renesse},
  {and} \bibinfo{person}{Emin~Gün Sirer}.} \bibinfo{year}{2018}\natexlab{}.
\newblock \bibinfo{booktitle}{\emph{Decentralization in Bitcoin and Ethereum
  Networks}}. \bibinfo{series}{Lecture Notes in Computer Science},
  Vol.~\bibinfo{volume}{10957}.
\newblock \bibinfo{publisher}{Springer Berlin Heidelberg},
  \bibinfo{address}{Berlin, Heidelberg}, \bibinfo{pages}{439–457}.
\newblock
\showISBNx{978-3-662-58386-9}
\urldef\tempurl%
\url{https://doi.org/10.1007/978-3-662-58387-6_24}
\showDOI{\tempurl}


\bibitem[Heimbach and Wattenhofer(2022)]%
        {heimbach2022eliminating}
\bibfield{author}{\bibinfo{person}{Lioba Heimbach} {and} \bibinfo{person}{Roger
  Wattenhofer}.} \bibinfo{year}{2022}\natexlab{}.
\newblock \showarticletitle{Eliminating Sandwich Attacks with the Help of Game
  Theory}.
\newblock \bibinfo{journal}{\emph{arXiv preprint arXiv:2202.03762}}
  (\bibinfo{year}{2022}).
\newblock


\bibitem[Juels et~al\mbox{.}(2021)]%
        {juels2021Apr}
\bibfield{author}{\bibinfo{person}{Ari Juels}, \bibinfo{person}{Ittay Eyal},
  {and} \bibinfo{person}{Mahimna Kelkar}.} \bibinfo{year}{2021}\natexlab{}.
\newblock \showarticletitle{{Miners, Front-Running-as-a-Service Is Theft}}.
\newblock \bibinfo{journal}{\emph{CoinDesk}} (\bibinfo{date}{Apr}
  \bibinfo{year}{2021}).
\newblock
\urldef\tempurl%
\url{https://www.coindesk.com/markets/2021/04/07/miners-front-running-as-a-service-is-theft}
\showURL{%
\tempurl}


\bibitem[Kelkar et~al\mbox{.}(2021)]%
        {kelkar2021themis}
\bibfield{author}{\bibinfo{person}{Mahimna Kelkar}, \bibinfo{person}{Soubhik
  Deb}, \bibinfo{person}{Sishan Long}, \bibinfo{person}{Ari Juels}, {and}
  \bibinfo{person}{Sreeram Kannan}.} \bibinfo{year}{2021}\natexlab{}.
\newblock \showarticletitle{Themis: Fast, Strong Order-Fairness in Byzantine
  Consensus}.
\newblock \bibinfo{journal}{\emph{Cryptology ePrint Archive}}
  (\bibinfo{year}{2021}).
\newblock


\bibitem[Kelkar et~al\mbox{.}(2020)]%
        {kelkar2020order}
\bibfield{author}{\bibinfo{person}{Mahimna Kelkar}, \bibinfo{person}{Fan
  Zhang}, \bibinfo{person}{Steven Goldfeder}, {and} \bibinfo{person}{Ari
  Juels}.} \bibinfo{year}{2020}\natexlab{}.
\newblock \showarticletitle{Order-fairness for byzantine consensus}. In
  \bibinfo{booktitle}{\emph{Annual International Cryptology Conference}}.
  Springer, \bibinfo{pages}{451--480}.
\newblock


\bibitem[Kiffer et~al\mbox{.}(2021)]%
        {kiffer2021gosssip}
\bibfield{author}{\bibinfo{person}{Lucianna Kiffer}, \bibinfo{person}{Asad
  Salman}, \bibinfo{person}{Dave Levin}, \bibinfo{person}{Alan Mislove}, {and}
  \bibinfo{person}{Cristina Nita-Rotaru}.} \bibinfo{year}{2021}\natexlab{}.
\newblock \showarticletitle{Under the Hood of the Ethereum Gossip Protocol}.
  \bibinfo{pages}{26}.
\newblock


\bibitem[Kokoris-Kogias et~al\mbox{.}(2018)]%
        {kokoris2018calypso}
\bibfield{author}{\bibinfo{person}{Eleftherios Kokoris-Kogias},
  \bibinfo{person}{Enis~Ceyhun Alp}, \bibinfo{person}{Linus Gasser},
  \bibinfo{person}{Philipp Jovanovic}, \bibinfo{person}{Ewa Syta}, {and}
  \bibinfo{person}{Bryan Ford}.} \bibinfo{year}{2018}\natexlab{}.
\newblock \showarticletitle{CALYPSO: Private data management for decentralized
  ledgers}.
\newblock \bibinfo{journal}{\emph{Cryptology ePrint Archive}}
  (\bibinfo{year}{2018}).
\newblock


\bibitem[Kursawe(2020)]%
        {kursawe2020wendy}
\bibfield{author}{\bibinfo{person}{Klaus Kursawe}.}
  \bibinfo{year}{2020}\natexlab{}.
\newblock \showarticletitle{Wendy, the good little fairness widget: Achieving
  order fairness for blockchains}. In \bibinfo{booktitle}{\emph{Proceedings of
  the 2nd ACM Conference on Advances in Financial Technologies}}.
  \bibinfo{pages}{25--36}.
\newblock


\bibitem[Leonardos et~al\mbox{.}(2021)]%
        {eip1559analysis}
\bibfield{author}{\bibinfo{person}{Stefanos Leonardos},
  \bibinfo{person}{Barnabé Monnot}, \bibinfo{person}{Daniël Reijsbergen},
  \bibinfo{person}{Stratis Skoulakis}, {and} \bibinfo{person}{Georgios
  Piliouras}.} \bibinfo{year}{2021}\natexlab{}.
\newblock \showarticletitle{Dynamical Analysis of the EIP-1559 Ethereum Fee
  Market}.
\newblock \bibinfo{journal}{\emph{arXiv:2102.10567 [cs, math]}}
  (\bibinfo{date}{Jun} \bibinfo{year}{2021}).
\newblock
\urldef\tempurl%
\url{http://arxiv.org/abs/2102.10567}
\showURL{%
\tempurl}
\newblock
\shownote{arXiv: 2102.10567}.


\bibitem[Lewis(2014)]%
        {flashboys}
\bibfield{author}{\bibinfo{person}{Michael Lewis}.}
  \bibinfo{year}{2014}\natexlab{}.
\newblock \bibinfo{booktitle}{\emph{Flash Boys}}.
\newblock \bibinfo{publisher}{W.W. Norton \& Company}.
\newblock
\showISBNx{978-0-393-24467-0}


\bibitem[Manahov(2016)]%
        {Manahov_2016}
\bibfield{author}{\bibinfo{person}{Viktor Manahov}.}
  \bibinfo{year}{2016}\natexlab{}.
\newblock \showarticletitle{Front-Running Scalping Strategies and Market
  Manipulation: Why Does High-Frequency Trading Need Stricter Regulation?}
\newblock \bibinfo{journal}{\emph{Financial Review}} \bibinfo{volume}{51},
  \bibinfo{number}{3} (\bibinfo{date}{Aug} \bibinfo{year}{2016}),
  \bibinfo{pages}{363–402}.
\newblock
\showISSN{07328516}
\urldef\tempurl%
\url{https://doi.org/10.1111/fire.12103}
\showDOI{\tempurl}


\bibitem[McMenamin et~al\mbox{.}(2022)]%
        {mcmenamin2022fairtradex}
\bibfield{author}{\bibinfo{person}{Conor McMenamin}, \bibinfo{person}{Vanesa
  Daza}, {and} \bibinfo{person}{Matthias Fitzi}.}
  \bibinfo{year}{2022}\natexlab{}.
\newblock \showarticletitle{FairTraDEX: A Decentralised Exchange Preventing
  Value Extraction}.
\newblock \bibinfo{journal}{\emph{arXiv preprint arXiv:2202.06384}}
  (\bibinfo{year}{2022}).
\newblock


\bibitem[Nakamoto(2008)]%
        {bitcoin}
\bibfield{author}{\bibinfo{person}{Satoshi Nakamoto}.}
  \bibinfo{year}{2008}\natexlab{}.
\newblock \showarticletitle{Bitcoin: A Peer-to-Peer Electronic Cash System}.
\newblock  (\bibinfo{year}{2008}), \bibinfo{pages}{9}.
\newblock


\bibitem[Piatt et~al\mbox{.}(2021)]%
        {eden}
\bibfield{author}{\bibinfo{person}{Chris Piatt}, \bibinfo{person}{Jeffrey
  Quesnelle}, {and} \bibinfo{person}{Caleb Sheridan}.}
  \bibinfo{year}{2021}\natexlab{}.
\newblock \showarticletitle{Eden Network}.
\newblock  (\bibinfo{year}{2021}).
\newblock
\urldef\tempurl%
\url{https://edennetwork.io/EDEN_Network___Whitepaper___2021_07.pdf}
\showURL{%
\tempurl}


\bibitem[Piet et~al\mbox{.}(2022)]%
        {pietExtractingGodlSic2022}
\bibfield{author}{\bibinfo{person}{Julien Piet}, \bibinfo{person}{Jaiden
  Fairoze}, {and} \bibinfo{person}{Nicholas Weaver}.}
  \bibinfo{year}{2022}\natexlab{}.
\newblock \showarticletitle{Extracting Godl [sic] from the Salt Mines: Ethereum
  Miners Extracting Value}.
\newblock \bibinfo{journal}{\emph{arXiv:2203.15930 [cs]}} (\bibinfo{date}{Mar}
  \bibinfo{year}{2022}).
\newblock
\urldef\tempurl%
\url{http://arxiv.org/abs/2203.15930}
\showURL{%
\tempurl}
\newblock
\shownote{arXiv: 2203.15930}.


\bibitem[Qin et~al\mbox{.}(2021)]%
        {qinQuantifyingBlockchainExtractable2021}
\bibfield{author}{\bibinfo{person}{Kaihua Qin}, \bibinfo{person}{Liyi Zhou},
  {and} \bibinfo{person}{Arthur Gervais}.} \bibinfo{year}{2021}\natexlab{}.
\newblock \showarticletitle{Quantifying Blockchain Extractable Value: How dark
  is the forest?}
\newblock \bibinfo{journal}{\emph{arXiv:2101.05511 [cs]}} (\bibinfo{date}{Dec}
  \bibinfo{year}{2021}).
\newblock
\urldef\tempurl%
\url{http://arxiv.org/abs/2101.05511}
\showURL{%
\tempurl}
\newblock
\shownote{arXiv: 2101.05511}.


\bibitem[Szabo(1997)]%
        {szabo1997}
\bibfield{author}{\bibinfo{person}{Nick Szabo}.}
  \bibinfo{year}{1997}\natexlab{}.
\newblock \bibinfo{title}{Formalizing and Securing Relationships on Public
  Networks}.
\newblock
\newblock
\urldef\tempurl%
\url{https://firstmonday.org/ojs/index.php/fm/article/download/548/469}
\showURL{%
\tempurl}


\bibitem[Torres et~al\mbox{.}(2021)]%
        {ferreira2021frontrunner}
\bibfield{author}{\bibinfo{person}{Christof~Ferreira Torres},
  \bibinfo{person}{Ramiro Camino}, {and} \bibinfo{person}{Radu State}.}
  \bibinfo{year}{2021}\natexlab{}.
\newblock \showarticletitle{Frontrunner Jones and the Raiders of the Dark
  Forest: An Empirical Study of Frontrunning on the Ethereum Blockchain}. In
  \bibinfo{booktitle}{\emph{USENIX Security Symposium, Virtual 11-13 August
  2021}}.
\newblock


\bibitem[Wang et~al\mbox{.}(2020)]%
        {wang2020flashloans}
\bibfield{author}{\bibinfo{person}{Dabao Wang}, \bibinfo{person}{Siwei Wu},
  \bibinfo{person}{Ziling Lin}, \bibinfo{person}{Lei Wu},
  \bibinfo{person}{Xingliang Yuan}, \bibinfo{person}{Yajin Zhou},
  \bibinfo{person}{Haoyu Wang}, {and} \bibinfo{person}{Kui Ren}.}
  \bibinfo{year}{2020}\natexlab{}.
\newblock \showarticletitle{Towards understanding flash loan and its
  applications in defi ecosystem}.
\newblock \bibinfo{journal}{\emph{CoRR}}  \bibinfo{volume}{abs/2010.12252}
  (\bibinfo{year}{2020}).
\newblock


\bibitem[Wood et~al\mbox{.}(2014)]%
        {ethereum}
\bibfield{author}{\bibinfo{person}{Gavin Wood} {et~al\mbox{.}}}
  \bibinfo{year}{2014}\natexlab{}.
\newblock \showarticletitle{Ethereum: A secure decentralised generalised
  transaction ledger}.
\newblock \bibinfo{journal}{\emph{Ethereum project yellow paper}}
  \bibinfo{volume}{151}, \bibinfo{number}{2014} (\bibinfo{year}{2014}),
  \bibinfo{pages}{1--32}.
\newblock


\bibitem[Yakira et~al\mbox{.}(2021)]%
        {yakira2021helix}
\bibfield{author}{\bibinfo{person}{David Yakira}, \bibinfo{person}{Avi Asayag},
  \bibinfo{person}{Gad Cohen}, \bibinfo{person}{Ido Grayevsky},
  \bibinfo{person}{Maya Leshkowitz}, \bibinfo{person}{Ori Rottenstreich}, {and}
  \bibinfo{person}{Ronen Tamari}.} \bibinfo{year}{2021}\natexlab{}.
\newblock \showarticletitle{Helix: A fair blockchain consensus protocol
  resistant to ordering manipulation}.
\newblock \bibinfo{journal}{\emph{IEEE Transactions on Network and Service
  Management}} \bibinfo{volume}{18}, \bibinfo{number}{2}
  (\bibinfo{year}{2021}), \bibinfo{pages}{1584--1597}.
\newblock


\bibitem[Zhou et~al\mbox{.}(2021a)]%
        {zhou2021a2mm}
\bibfield{author}{\bibinfo{person}{Liyi Zhou}, \bibinfo{person}{Kaihua Qin},
  {and} \bibinfo{person}{Arthur Gervais}.} \bibinfo{year}{2021}\natexlab{a}.
\newblock \showarticletitle{A2mm: Mitigating frontrunning, transaction
  reordering and consensus instability in decentralized exchanges}.
\newblock \bibinfo{journal}{\emph{arXiv preprint arXiv:2106.07371}}
  (\bibinfo{year}{2021}).
\newblock


\bibitem[Zhou et~al\mbox{.}(2021b)]%
        {zhouHighFrequency2021}
\bibfield{author}{\bibinfo{person}{Liyi Zhou}, \bibinfo{person}{Kaihua Qin},
  \bibinfo{person}{Christof~Ferreira Torres}, \bibinfo{person}{Duc~Viet Le},
  {and} \bibinfo{person}{Arthur Gervais}.} \bibinfo{year}{2021}\natexlab{b}.
\newblock \showarticletitle{High-Frequency Trading on Decentralized On-Chain
  Exchanges}. In \bibinfo{booktitle}{\emph{42nd {IEEE} Symposium on Security
  and Privacy, {SP} 2021, San Francisco, CA, USA, 24-27 May 2021}}.
  \bibinfo{publisher}{{IEEE}}, \bibinfo{pages}{428--445}.
\newblock


\end{thebibliography}

\end{document}